\documentclass[aps,prx,reprint,showpacs,amsmath,amssymb,floatfix,superscriptaddress,noeprint]{revtex4-2}

\usepackage[colorlinks,linkcolor=blue,urlcolor=blue,citecolor=blue,anchorcolor=blue]{hyperref}

\usepackage{graphicx} 

\usepackage{physics}
\usepackage{siunitx}
\DeclareSIUnit{\belmilliwatt}{Bm}
\DeclareSIUnit{\dBm}{\deci\belmilliwatt}
\usepackage{tensor}
\usepackage{ulem}
\usepackage{comment}
\usepackage{xcolor}
\usepackage{dsfont}
\usepackage{makecell}

\begin{document}
\title{Autoparametric resonance extending the bit-flip time of a cat qubit up to 0.3~s}

\author{A. Marquet}
\affiliation{Alice \& Bob, 49 Bd du G\'en\'eral Martial Valin, 75015 Paris, France}
\affiliation{Ecole Normale Sup\'erieure de Lyon,  CNRS, Laboratoire de Physique, F-69342 Lyon, France}
\author{A. Essig}
\affiliation{Alice \& Bob, 49 Bd du G\'en\'eral Martial Valin, 75015 Paris, France}
\author{J. Cohen}
\affiliation{Alice \& Bob, 49 Bd du G\'en\'eral Martial Valin, 75015 Paris, France}
\author{N. Cottet}
\affiliation{Alice \& Bob, 49 Bd du G\'en\'eral Martial Valin, 75015 Paris, France}
\author{A. Murani}
\affiliation{Alice \& Bob, 49 Bd du G\'en\'eral Martial Valin, 75015 Paris, France}
\author{E. Albertinale}
\affiliation{Alice \& Bob, 49 Bd du G\'en\'eral Martial Valin, 75015 Paris, France}
\author{S. Dupouy}
\affiliation{Alice \& Bob, 49 Bd du G\'en\'eral Martial Valin, 75015 Paris, France}
\affiliation{Ecole Normale Sup\'erieure de Lyon,  CNRS, Laboratoire de Physique, F-69342 Lyon, France}
\author{A. Bienfait}
\affiliation{Ecole Normale Sup\'erieure de Lyon,  CNRS, Laboratoire de Physique, F-69342 Lyon, France}
\author{T. Peronnin}
\affiliation{Alice \& Bob, 49 Bd du G\'en\'eral Martial Valin, 75015 Paris, France}
\author{S. Jezouin}
\affiliation{Alice \& Bob, 49 Bd du G\'en\'eral Martial Valin, 75015 Paris, France}
\author{R. Lescanne}
\thanks{These authors co-supervised the project}
\affiliation{Alice \& Bob, 49 Bd du G\'en\'eral Martial Valin, 75015 Paris, France}
\author{B. Huard}
\thanks{These authors co-supervised the project}
\affiliation{Ecole Normale Sup\'erieure de Lyon,  CNRS, Laboratoire de Physique, F-69342 Lyon, France}

\date{\today}

\begin{abstract}
Cat qubits, for which logical $|0\rangle$ and $|1\rangle$ are coherent states $|\pm\alpha\rangle$ of a harmonic mode, offer a promising route towards quantum error correction. Using dissipation to our advantage so that photon pairs of the harmonic mode are exchanged with single photons of its environment, it is possible to stabilize the logical states and exponentially increase the bit-flip time of the cat qubit with the photon number $|\alpha|^2$. Large two-photon dissipation rate $\kappa_2$ ensures fast qubit manipulation and short error correction cycles, which are instrumental to correct the remaining phase-flip errors in a repetition code of cat qubits. Here we introduce and operate an autoparametric superconducting circuit that couples a mode containing the cat qubit to a lossy mode whose frequency is set at twice that of the cat mode. This passive coupling does not require a parametric pump and reaches a rate $\kappa_2/2\pi\approx 2~\mathrm{MHz}$. With such a strong two-photon dissipation, bit-flip errors of the autoparametric cat qubit are prevented for a characteristic time up to 0.3~s with only a mild impact on phase-flip errors. Besides, we illustrate how the phase of a quantum superposition between $|\alpha\rangle$ and $|-\alpha\rangle$ can be arbitrarily changed by driving the harmonic mode while keeping the engineered dissipation active.
\end{abstract}
\maketitle

Quantum error correction is instrumental in building useful quantum processors. It is based on the gathering of many physical quantum systems in order to form protected logical qubits. The number of required physical systems can be daunting but strategies involving harmonic oscillators instead of qubits promise to reduce that number by a large factor~\cite{Mirrahimi2014,Cai2021,Joshi2021,Ma2021,Albert2022,Guillaud2023}.
\begin{figure}[h!]
\includegraphics[width=\linewidth]{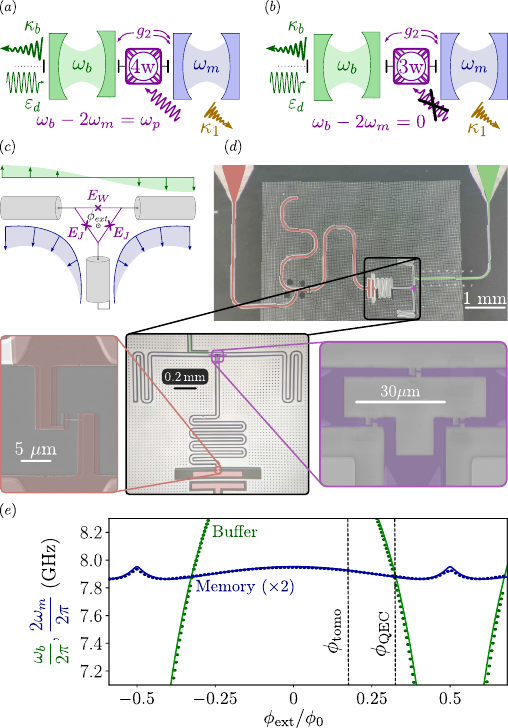}
\caption{(a) A four-wave mixing coupler such as a transmon or ATS swaps pairs of photons of a memory mode at $\omega_m$ for single photons of a buffer mode at $\omega_b$ at a rate $g_2$ owing to a pump at $|\omega_b-2\omega_m|$. The buffer loss rate $\kappa_b$ thus leads to an effective two-photon dissipation rate $\kappa_2$ that scales with pump amplitude. Driving the buffer mode on resonance at a displacement rate $\epsilon_d$ stabilizes a cat code $\{|\pm\alpha\rangle\}$. (b) A three-wave mixing coupler passively performs the same photon exchange when $\omega_b=2\omega_m$.
(c) Scheme of the autoparametric cat. The three-wave mixing coupler is a ring of three Josephson junctions threaded by a flux $\phi_\mathrm{ext} = \varphi_\mathrm{ext}\varphi_0$, 
with $\varphi_0=\hbar/2e$, Josephson energies $E_W/h\approx 115~\mathrm{GHz}$ and $E_J/h\approx 250~\mathrm{GHz}$. The buffer (green) and memory (blue) mode geometries are represented as field vectors. (d) Optical and e-beam images of a clone of the device. False colors highlight the input buffer/flux line (green), the tomography transmon and its readout resonator and Purcell filter (red) as well as the three-wave mixing coupler (purple) (see Sec.~\ref{Appendix:Cabling}). An array of holes in the tantalum (to prevent the apparition of vortices) appear as a white rectangle in the larger image. (e) Dots: measured resonance frequency of the buffer $\omega_b$ (green) and twice the measured resonance frequency of the memory $2\omega_m$ (blue) as a function of the flux threading the ring. Dashed lines indicate the flux biases where the circuit is operated. Solid lines: Best fit of the circuit parameters (see Sec.~\ref{Appendix:Hamiltonian_derivation}).}
\label{fig:fig1}
\end{figure}
A prominent example is the cat qubit, whose computational states are coherent states $|\pm\alpha\rangle$ of a harmonic oscillator, such as a superconducting microwave cavity~\cite{PhysRevA.65.042305,Ralph2003,Jeong2007,Mirrahimi2014}. These states can be stabilized through measurement based feedback ~\cite{Ofek2016,Rosenblum2018,Grimsmo2020}, Hamiltonian engineering~\cite{Puri2017, Grimm2020, Frattini2022,Iyama2023} or reservoir engineering~\cite{Leghtas2015, Touzard2018, Lescanne2020exponential,Putterman2022,Gravina2023,Gertler2023}. The latter strategy prevents the state of that \textit{memory} cavity from leaking out of the cat qubit subspace by engineering its coupling to the environment with the key feature that photon pairs are lost at a rate $\kappa_2$. The bit flip time $T_\mathrm{X}$ then increases exponentially with the photon number $|\alpha|^2$ at the modest expense of a linear deterioration of the phase flip rate $\Gamma_\mathrm{Z}\propto|\alpha|^2$~\cite{Lescanne2020exponential}. Reaching large values of two-photon rate $\kappa_2$ is critical in this strategy. First, to ensure exponential bit-flip protection, it should overcome any parasitic processes affecting the memory such as dephasing, thermal excitation, Kerr effect, frequency shifts due to a thermally populated ancilla qubit~\cite{Lescanne2020exponential, Berdou2022} or gate drives. Second, its value sets a higher bound on cat qubit gate speed and needs to be large compared to the residual single photon loss rate $\kappa_1$, the main cause of phase-flip errors~\cite{LeRegent2022}. The remaining phase-flip errors could then be corrected using a repetition code made of a chain of cat qubits under the condition $\kappa_2 / \kappa_1 \gtrsim  10^2$~\cite{Chamberland2022, Guillaud2019, LeRegent2022}. Introducing a new non-linear design called the \textit{autoparametric cat}, our experiment benefits from a stronger 3-wave mixing interaction compared to the 4-wave mixing interaction of previous schemes. We demonstrate $\kappa_2$ rates as high as $2\times 2\pi~\mathrm{MHz}$, about 150 times larger than $\kappa_1$. As we increase the photon number, we observe an improvement of the bit-flip time by more than a factor $25,000$, reaching $0.36\pm 0.15~\mathrm{s}$ while the phase-flip rate degrades by less than a factor of $6$. In addition, we demonstrate a Z gate of the cat code with a fidelity of $96.5\pm 2\%$ in 28~ns.

To achieve two-photon coupling between the memory and its environment, an intermediary mode with single photon coupling $\kappa_b$ to the environment acts as a buffer and a two-to-one photon exchange Hamiltonian
$\hat{H}=\hbar g_2 \hat{m}^2 \hat{b}^\dagger +h.c.$ is activated with a rate $g_2$. The operators $\hat{m}$ and $\hat{b}$ are the annihilation operators of the memory and buffer mode respectively. In the limit where $g_2$ is small enough compared to $\kappa_b$, the two-photon dissipation rate reads $\kappa_2 = 4g_2^2/\kappa_b$. This interaction has been engineered in the past by parametrically pumping a 4 wave mixing non-linearity (Fig.~\ref{fig:fig1}a), such as a single junction or an asymmetrically threaded squid (ATS), at the frequency matching condition $\omega_p = 2\omega_m - \omega_b$. However, although the two-to-one photon exchange rate $g_2$ scales with the pump amplitude, increasingly large pump powers are known to affect coherence times and activate higher non-linear terms in the Hamiltonian~\cite{Touzard2018}. In practice, the limited range of $\kappa_2$ rates that could be obtained with pumped nonlinearities prevents surpassing the self-Kerr rate or the dispersive coupling rate between transmon and memory~\cite{Lescanne2020exponential,Touzard2018,Gautier2022}. In this work, we use 3 wave mixing instead (Fig.~\ref{fig:fig1}b), thus alleviating the need for a pump to mediate the 2 photon interaction as in the 1989 proposal in Ref.~\cite{Wolinsky1988}. The frequency matching condition then becomes $2\omega_m = \omega_b$. 
This condition is characteristic of autoparametric systems so that the buffer field passively performs a parametric driving of the memory~\cite{Verhulst2002}. Remarkably, the resulting exchange rate $g_2$ is much larger than what can be reached for 4 wave mixing (see comparison in section~\ref{Appendix:comparison}).

The mixing element of the autoparametric cat consists of two main Josephson junctions with energy $E_J$ symmetrically arranged within a superconducting loop that is threaded with an external magnetic flux $\phi_\mathrm{ext}$ (Fig.~\ref{fig:fig1}c). These two junctions in parallel configuration have a common mode serving as a memory mode and a differential mode, associated with the flux degree of freedom of the loop and serving as a buffer mode (see Sec.~\ref{Appendix:circuit_design}). In order to lower the relatively high frequency of the buffer mode and increase its flux tunability, a third Josephson junction with energy $E_W$ is added in the loop making its configuration similar to previously realized circuits~\cite{Vion2002,PhysRevLett.115.203601,PhysRevB.93.134501,Dassonneville2019,Lu2023}. Besides, it endows the memory mode with a frequency sweet spot provided $E_W < E_J/\sqrt{2}$. By symmetry, the memory mode does not participate in this third junction which plays no role in the two-to-one photon exchange Hamiltonian. To tune the mode frequencies and participation in the mixing element, the superconducting loop is further integrated within a linear microwave network that preserves the mode symmetries  (Fig.~\ref{fig:fig1}c). A single input line (green in Fig.~\ref{fig:fig1}d) then couples to the circuit in order to provide both fast flux bias and drive the buffer. The frequency tunability makes it difficult to engineer a filter that protects the memory lifetime. Instead, we leverage the symmetries of the circuit (Fig.~\ref{fig:fig1}d) and position the input line such that it does not impact the memory quality factor while preserving a strong coupling to the buffer (see Sec.~\ref{Appendix:Circuit_design}). We achieve a  buffer coupling rate $\kappa_b/2\pi \approx 40~\mathrm{MHz}$ and a much lower memory loss rate $\kappa_1/2\pi \approx 14~\mathrm{kHz}$. Note that an RF-SQUID or a SNAIL~\cite{Frattini2017} could have been used as a 3-wave mixing element. However, the existence of a sweet-spot in flux and the possibility to leverage the circuit symmetries to preserve the memory quality factor made us favor this design. Finally, a transmon qubit is inductively coupled (see section~\ref{sec:ind_coupling}) to the memory with $\chi/2\pi = 170~\mathrm{kHz}$ to perform the Wigner tomography of the memory mode ~\cite{Davidovich1996,Bertet2002,Vlastakis2013}. Note that a work conducted in parallel to ours demonstrates that Wigner tomography can be performed without the transmon~\cite{Reglade2023}.

The mixing element enforces a two-to-one photon exchange Hamiltonian with strength $g_2$. Around the memory sweet spot, its value is well approximated by (see Sec.~\ref{Appendix:Hamiltonian_derivation}) 
\begin{equation}
\label{eq:g2}
 g_2 \approx \frac{E_W}{\hbar}\left(1-\frac{{\delta\varphi_\mathrm{ext}}^2}{2}\right)\varphi_{\mathrm{zpf},m}^2 \varphi_{\mathrm{zpf},b}
\end{equation}
where $\delta\varphi_\mathrm{ext}=(\phi_\mathrm{ext}-\phi_\mathrm{ext}^{(\mathrm{sweet})})/\varphi_0$ is the distance from the sweet spot and $\varphi_{\mathrm{zpf},m/b}$ are the zero point fluctuations of the phase from the memory and buffer mode respectively across either of the main junction. In practice, it is hard to ensure the frequency matching condition precisely at the sweet spot. In the experiment, $g_2$ is close to its maximum value at the flux $\phi_\mathrm{QEC}$ for which the frequency of the buffer matches twice the frequency of the memory $\omega_b(\phi_\mathrm{QEC})=2\omega_m(\phi_\mathrm{QEC})\approx ~2\pi\times 7.896 \mathrm{GHz}$ (see Fig.~\ref{fig:fig1}e). Therefore the three-wave mixing term performs the desired swap between pairs of memory photons and single buffer photons at $\phi_\mathrm{QEC}$. 

\begin{figure}[h!]
\includegraphics[width=\linewidth]{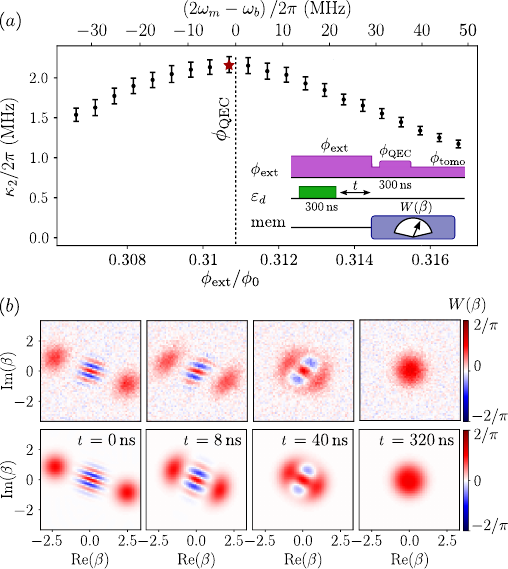}
\caption{(a) Dots: Measured two-photon relaxation rate $\kappa_2$ as a function of flux bias close to $\phi_\mathrm{QEC}$. Error bars represent statistical uncertainties. Inset: pulse sequence used for the measurement. The detuning between buffer frequency and twice the memory frequency is indicated on the top axis. (b) Top: Measured Wigner functions of the memory after the decay times indicated on the figure for $\left|\alpha\right|=2.5$ and at the flux indicated by the star in (a). Bottom: results of the simulation using $\kappa_2/2\pi=2.16~\mathrm{MHz}$. The buffer mode is adiabatically eliminated in these simulations to provide an effective value of $\kappa_2$ (see Sec.~\ref{Appendix:fit_kappas}).}
\label{fig:fig2}
\end{figure} 

The preparation of a cat state $|C_+^\alpha\rangle\propto|\alpha\rangle+|-\alpha\rangle$ consists in starting from the vacuum state in the memory at $\phi_\mathrm{QEC}$ and turning on a drive with an amplitude $|\epsilon_d|=|\alpha|^2 g_2$ (Fig.~\ref{fig:fig1}b) at twice the memory frequency $\omega_d=2\omega_m$ for a time of $300$~ns. It is shorter than the characteristic time  $1/|\alpha|^2\kappa_1$ of a photon loss yet long enough for the two-photon dissipation and drive to stabilize the memory into the cat qubit manifold. From there, one may determine the two-photon loss rate $\kappa_2$ by monitoring how the memory decays once the drive has been turned off. Fitting the measured Wigner function $W(\beta)$ at various time steps with the solution of a Lindblad master equation where $\kappa_2$ is the single free parameter allows us to determine $\kappa_2/2\pi\approx 2.16\pm 0.1~\mathrm{MHz}$ at $\phi_\mathrm{QEC}$. The ratio $\kappa_2/\kappa_1\approx 1.5\times 10^2$ is much larger than in previous implementations of two-photon dissipation using four-wave mixing~\cite{Leghtas2015,Touzard2018,lescanne2020,Berdou2022}. This procedure is repeated for various flux biases around $\phi_\mathrm{QEC}$ in Fig.~\ref{fig:fig2}a, which shows the range of detuning $\omega_b-2\omega_m$ over which the two-photon loss rate decreases. The discrepancy between experiment and simulation of the evolution of the Wigner functions (Fig.~\ref{fig:fig2}b), mainly visible at 8~ns, can be attributed to the breakdown of the condition for adiabatic elimination of the buffer mode (see Sec.~\ref{Appendix:fit_kappas}). Indeed this approximation breaks down when the frequency $8 g_2 |\alpha|$ at which single buffer photons are swapped with pairs of memory photons gets larger than the rate $\kappa_b$ at which they leak out from the buffer.

Note that Wigner tomography of $W(\beta)$ is performed using a displacement of the memory mode by $\hat{D}(-\beta)$, followed by a parity measurement that exploits the dispersively coupled qubit in a Ramsey-like sequence~Ref.~\cite{Davidovich1996,Bertet2002,Vlastakis2013}. However, since both displacements and the dispersive coupling are inhibited by two-photon loss, these operations are performed at a flux $\phi_\mathrm{tomo}$ (see Fig.~\ref{fig:fig1}e) such that the strong detuning $\omega_b-2\omega_m$ disables two-photon dissipation, yet without too much nonlinearity that would distort the tomography. In practice, the tomography thus requires to abruptly change the flux bias between $\phi_\mathrm{QEC}$ and $\phi_\mathrm{tomo}$ (see Sec.~\ref{Appendix:Cat_Wigner_phase_correction}). Besides, we improve the performance of this tomography by initializing the parity measurement with an idle time of 300~ns at $\phi_\mathrm{QEC}$ so that all pairs of photons are dumped into the environment. The parity measurement then boils down to determining whether or not a single photon remains in the memory~\cite{Chamberland2022, Marquet2023bis}. Parity is preserved during this operation owing to the large ratio $\kappa_2/\kappa_1$.

\begin{figure}[h!]
\includegraphics[width=\linewidth]{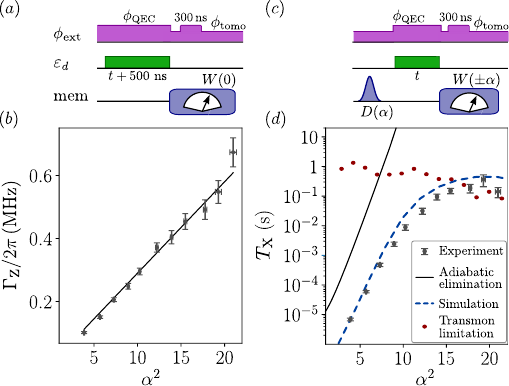}
\caption{(a) Pulse sequence of the phase flip rate measurement. (b) Dots: measured phase flip rate $\Gamma_{Z}$ (linear scale) of the cat code as a function of photon number $|\alpha|^2$. All values are obtained by fitting $W(0)$ to an exponential decay in time. Error bars represent statistical uncertainties (see Sec.~\ref{Appendix:Exemples_Tbf_Gammapf}). Line: expected rate $2|\alpha|^2\kappa_1$. (c) Pulse sequence of the bit-flip time measurement. (d) Dots: measured bit flip time (log scale) of the cat code as a function of photon number $|\alpha|^2$. All values are obtained by fitting the difference $W(\alpha)-W(-\alpha)$ to an exponential decay in time. Error bars represent statistical uncertainties (see Sec.~\ref{Appendix:Exemples_Tbf_Gammapf}). Solid black line: expected bit-flip time with $\kappa_\varphi^{m}/2\pi=0.16~\mathrm{MHz}$ under the adiabatic elimination of the buffer: $e^{2|\alpha|^2}/(|\alpha|^2\kappa_\varphi^{m})$. Dashed blue line: simulated bit-flip time with the same $\kappa_\varphi^{m}$, assuming a detuning $2\omega_m-\omega_b=2\pi\times 3.5~\mathrm{MHz}$. Red dots: bound below which $T_\mathrm{X}$ is not limited by excitation of higher states of the transmon~(Fig.~\ref{fig:Thermal_pop}).
}
\label{fig:figrates}
\end{figure}

The phase-flip rate $\Gamma_{Z}$ of the cat code can be measured in a similar manner as the two photon loss rate (Fig.~\ref{fig:figrates}a). Starting from the quantum superposition $|C_+^{\alpha}\rangle$, the parity, hence the measured $W(0)$, decays exponentially with time at a rate $\Gamma_{Z}$ while the drive $\epsilon_d(\alpha)$ is kept on. In Fig.~\ref{fig:figrates}b are shown the observed $\Gamma_{Z}$ rates as a function of photon number $\left| \alpha \right|^2$ in the cat code. As expected, the phase-flip rate increases linearly as $\Gamma_{Z}=2|\alpha|^2\kappa_1$ until it goes above 20 photons.

\begin{figure}[h!]
\includegraphics[width=\linewidth]{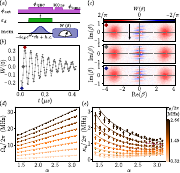}
\caption{(a) Pulse sequence used to perform a $Z$ rotation in the cat qubit $\{|\alpha\rangle,|-\alpha\rangle\}$ basis. (b) Dots: Measured oscillations of $W(0)$ as a function of time $t$ using the pulse sequence of Fig.~\ref{fig:figrates}a where the cat code is stabilized with a photon number $|\alpha|^2=9.3$. An additional displacement drive at $\omega_m$ starts $240~\mathrm{ns}$ after the buffer drive is turned on. Here, its amplitude $\epsilon_Z(t)$ is Gaussian shaped with a mean amplitude $\bar{\epsilon}_Z/2\pi=1.625~\mathrm{MHz}$ and its phase is chosen to displace in the direction indicated by the arrow in c). Line: fit to oscillations at a frequency $\Omega_Z/2\pi=19.8~\mathrm{MHz}$, which are decaying at a rate $\kappa_Z/2\pi=0.62~\mathrm{MHz}$. (c) Measured Wigner functions $W(\beta)$ after a $Z$ rotation of angle $\theta=2\pi$, $3\pi/2$, and $\pi$ from top to bottom.
(d) Dots: Inferred rotation frequency $\Omega_Z$ as a function of cat code amplitude $\alpha$, and for various mean drive amplitudes $\bar{\epsilon}_Z/2\pi=0.32,0.625,0.965,1.295,1.625,1.955,2.285$, and $2.66~\mathrm{MHz}$ from bright to dark orange. Lines: expected rotation frequency $\Omega_Z=4\mathrm{Re}(\bar{\epsilon}_Z\alpha)$ around $Z$. (e) Dots: Inferred decay rate $\kappa_Z$ as a function of $|\alpha|$ for the same drive amplitudes. Lines: simulated decay rate with $g_2/2\pi=6~\mathrm{MHz}$ as a fit parameter and the same detuning as in Fig.~\ref{fig:figrates}d.}
\label{fig:Zoscillations}
\end{figure}

The bit-flip time $T_\mathrm{X}$ characterizes how fast the coherent state $|\alpha\rangle$ decays to an equal mixture of $|-\alpha\rangle$ and $|\alpha\rangle$. In order to measure it, the flux is first set to $\phi_\mathrm{tomo}$ so that a memory drive can prepare the state $|\alpha\rangle$. The flux is then turned back to $\phi_\mathrm{QEC}$ and the buffer drive is immediately turned on with an amplitude $\epsilon_d(\alpha)$ (Fig.~\ref{fig:figrates}c). We measure the Wigner functions $W(\pm \alpha)$ for various waiting times and fit their difference $W(\alpha)-W(-\alpha)\propto e^{-t/T_\mathrm{X}}$. The resulting bit-flip time is shown in Fig.~\ref{fig:figrates}d and rises exponentially with photon number $|\alpha|^2$ until about 12 photons. There, $T_\mathrm{X}$ grows by a scaling factor of about $3.5$ per added photon, smaller than the limit of $7.4$ predicted in case of pure dephasing alone (solid line)~\cite{LeRegent2022}.

We explain this discrepancy by the breakdown of the approximation of adiabatic elimination of the buffer. We performed simulations of the evolution of the buffer-memory bipartite system that indeed predict smaller scaling factors (dashed line in Fig.~\ref{fig:figrates}d and section~\ref{Appendix:Simus}). Some parameters we use in the master equation (\ref{eq:ME}) are experimentally measured such as the pure dephasing rate of the memory $\kappa_\varphi^{m}/2\pi = 0.16~\mathrm{MHz}$ (Fig.~\ref{fig:Memory_01_Ramsey}), the self-Kerr rate of the memory $\chi_{m,m}/2\pi = 0.22~\mathrm{MHz}$, and the measured detuning $\Delta_m/2\pi=3~\mathrm{MHz}$ between half the drive frequency and $\omega_m$. Other parameters are inferred from the circuit model and the pure dephasing rate of the buffer $\kappa_\varphi^{b}/2\pi$ is assumed to be limited by flux noise like $\kappa_\varphi^{m}/2\pi$ (see section~\ref{Appendix:Simus}). Besides the simulated bit-flip times strongly depend on the flux $\phi_\mathrm{ext}$ and we set a detuning $2\omega_m-\omega_b=2\pi\times 3.5~\mathrm{MHz}$ in order to better reproduce the measured bit-flip times. This corresponds to a flux offset of $3~\times 10^{-4}\phi_0$, which could be attributed to flux drifts during the month that separates the measurements of Fig.~\ref{fig:fig2} and \ref{fig:figrates}.

The bit-flip time saturates at $0.36\pm 0.15~\mathrm{s}$, reached for $|\alpha|^2\approx 20$. We identify three possible mechanisms that present comparable contributions to the bit-flip rate at large photon numbers. First, varying the dephasing rates $\kappa_\varphi^{m}$ in simulations lead to an apparent saturation of $T_\mathrm{X}$ (see Fig.~\ref{fig:Tx_vs_deltab_simulation}). A better tuning of the circuit parameters would cancel $\kappa_\varphi^{m}$ by matching the flux at which the autoparametric condition occurs with the memory sweet spot in Fig.~\ref{fig:fig1}e. A second candidate for the bit-flip limitation is the thermalization of the buffer mode, which may come from various origins (see section~\ref{Appendix:Bit_flip_time_limits}). In turn, owing to the breakdown of adiabatic elimination, it limits the bit-flip time. A mitigation strategy would consist in adapting the drive frequency for each size $\alpha$ of the cat qubit to keep $\Delta_m=0$. Finally, simulations reveal that while the transmon first excited state no longer limits $T_\mathrm{X}$~\cite{Lescanne2020exponential}, residual excitations of the transmon's higher states sets a bound below which the bit flip time cannot be affected by the transmon. Despite a relatively small dispersive shift $\chi/2\pi=170~\mathrm{kHz}$, these higher excited states of the transmon exit the dispersive regime yielding a large frequency shift on the memory comparable to $\kappa_2|\alpha|^2$, effectively turning off the cat qubit stabilization and inducing a bit-flip error. We measure the higher excited states of the transmon while stabilizing cat qubits of various mean photon number $|\alpha|^2$ (Fig.~\ref{fig:Thermal_pop}), and infer the rate at which they get populated (red dots in Fig.~\ref{fig:figrates}d). The measured saturation in bit-flip time seems to reach this bound, indicating that the excitation of transmon higher states may be a limiting mechanism. This limitation could be avoided by removing the transmon qubit altogether~\cite{Reglade2023}. 

With such a large ratio $\kappa_2/\kappa_1$, it is possible to realize $Z$ gates in the cat code $\{|\alpha\rangle,|-\alpha\rangle\}$ using quantum Zeno dynamics~\cite{Raimond2010,Raimond2012,Schafer2014,Signoles2014,Bretheau2015,Touzard2018}. Starting from $|C_+^{\alpha}\rangle$ with the buffer drive $\epsilon_d(\alpha)$ turned on, we continuously drive the memory on resonance with an amplitude $\epsilon_Z$~\cite{Touzard2018} (Fig.~\ref{fig:Zoscillations}a). The phase of the latter is chosen so that, without two-photon loss, the memory drive would induce a displacement (arrow in Fig.~\ref{fig:Zoscillations}c) perpendicular to $\alpha$ in the memory phase space~(see calibration procedure in Sec.~\ref{Appendix_Z_gate_calib}). The measured Wigner functions of the memory are shown in Fig.~\ref{fig:Zoscillations}c after three different waiting times. Owing to quantum Zeno dynamics, the combined effects of two-photon loss and buffer drive keep the memory state into the qubit space generated by $\{|\alpha\rangle,|-\alpha\rangle\}$, but the memory drive $\epsilon_Z$ rotates the phase $\theta$ of the quantum superposition $(|\alpha\rangle+e^{i\theta}|-\alpha\rangle)/\sqrt{\mathcal{N}}$, which can be seen as translated fringes in the Wigner function. The measured Wigner function $W(0)$ as a function of time in Fig.~\ref{fig:Zoscillations}b exhibits decaying oscillations around the $Z$ axis of the cat qubit at a frequency $\Omega_Z$ and decay rate $\kappa_Z$.

The rotation frequency is expected to be given by $\Omega_Z=4\mathrm{Re}(\epsilon_Z\alpha)$, which is precisely what we observe in Fig.~\ref{fig:Zoscillations}d. The decay rate $\kappa_Z$ has a more subtle dependence on photon number $|\alpha|^2$ and memory drive $\epsilon_Z$ as seen in Fig.~\ref{fig:Zoscillations}e. In the ideal case and for constant $\epsilon_Z$, the rate is expected to decay as $\kappa_Z=2\kappa_1|\alpha|^2+\kappa_b\epsilon_Z^2/(2|\alpha|^2g_2^2)$~\cite{Chamberland2022}. The first term corresponds to phase flips at a rate $\Gamma_{Z}$ while the second one corresponds to an induced leakage out of the confined cat qubit Hilbert space when the drive amplitude $\epsilon_Z$ is too strong to be Zeno blocked. This expression remains valid even outside of the adiabatic elimination regime~\cite{Gautier2023}.

In practice, the experiment deviates from this simple picture owing to the self-Kerr effect on the memory, slight detuning of the drive frequency, and resonance frequency detuning induced by drifts in the flux bias. Complete simulations of $\kappa_Z \left( \alpha\right)$ are sensitive to the two-to-one photon coupling rate $g_2$ so that they are used to determine it. Adjusting this parameter to $g_2/2\pi\approx 6\pm 0.5~\mathrm{MHz}$ leads to a good match between measurement and simulations (see Fig.~\ref{fig:Zoscillations}e). With such a coupling, the adiabatic elimination of the buffer predicts a larger two-photon dissipation rate $4g_2^2/\kappa_b\approx 3.6\times 2\pi~\mathrm{MHz}$ than what is measured in Fig.~\ref{fig:fig2}, which is expected since the condition $8 g_2 |\alpha| < \kappa_b$ is not met for $\left| \alpha \right|^2 \gtrsim 1$ ~\cite{Reglade2023}. Interestingly, despite the observed limitation on $T_\mathrm{X}$ in the non adiabatic regime, it is still possible to improve gate speed and fidelity by going to large values of $g_2|\alpha|$.

It is possible to infer the $Z$ gate fidelity from the measured evolution of $W(\beta)$ during the drive. Owing to the noise bias of the cat code, the $Z$ gate is the least faithful when applied to Clifford states $|\pm X\rangle$ and $|\pm Y\rangle$. In contrast, we measure that $|\pm Z\rangle$ states are mapped onto themselves by the $Z$ gate with less than $3\times 10^{-6}$ error probability under the driving conditions of Fig.~\ref{fig:Zoscillations}b (see Sec.~\ref{appendix:bias_gate}). The gate fidelity can be estimated to $F = 1/2+ \exp(-\pi \kappa_z/\Omega_z)/2$ (see Sec.~\ref{sec:gatefidelity}). Using Gaussian pulses for the memory drive leads to $Z$ gate fidelities $F=95\pm 2\%$ in 26~ns for $\left| \alpha \right|^2 = 9.3$ and for a drive amplitude $\bar{\epsilon}_Z/2\pi=1.625~\mathrm{MHz}$. 
We further improve the gate fidelity by using square pulses. Indeed, for a fixed average drive amplitude $\bar{\epsilon}_Z$, the square pulse has the smallest maximum amplitude, hence induces the least non-adiabatic errors. We then reach $F=96.5\pm 2\%$ in 28~ns for the $Z$ gate.

In this work, we show that the autoparametric approach to two-photon exchange can significantly enhance the two-photon dissipation rate $\kappa_2$, exceeding the values obtained with prior parametric pumping strategies. Remarkably, the autoparametric scheme does not seem to activate extra relaxation processes as indicated by the linear increase of phase-flip rate up to 20 average photons. We emphasize that repetition codes are more lenient with errors than surface codes~\cite{Ai2023}, which is another benefit of using strongly noise-biased qubits. Moreover, the associated photon exchange rate $g_2$ allows us to perform $Z$ gates with up to 96.5~\% fidelity in 28~ns. We achieved a notable bit flip time $T_\mathrm{X}$ of up to 0.3~s. Several mechanisms have been identified as possible limitations and mitigation strategies have been proposed for future realizations. The figure-of-merit $\kappa_2/\kappa_1 = 1.5\times 10^2$ exceeds the error correction threshold for a repetition code based on cat qubits~\cite{LeRegent2022}. We foresee that enhancing the memory lifetime by 1 or 2 orders of magnitude, for instance using a notch filter at the memory frequency and improving the fabrication process, could dramatically decrease the phase flip rate and bolster the scaling of error correction. Besides the present design parameters are quite conservative for this first demonstration. The zero point fluctuations of the modes in the junctions ($\varphi_{\mathrm{zpf}, m}= 0.0305 $ and $\varphi_{\mathrm{zpf}, b} = 0.0648$) can safely be increased, which would improve $g_2$. Additionally, in the future, the large coupling rates of the autoparametric approach could be leveraged under the frequency matching condition $\omega_b=4\omega_m$. This could then stabilize cat qubits that are composed of superpositions of the four coherent states $|\alpha\rangle$, $|i\alpha\rangle$, $|-\alpha\rangle$, and $|-i\alpha\rangle$, whose autonomous stabilization is still missing despite the strong interest of this approach for quantum error correction~\cite{Mirrahimi2014}.

\begin{acknowledgments} 
This research was supported by the QuantERA grant QuCos ANR-19-QUAN-0006, the Plan France 2030 through the project ANR-22-PETQ-0003. We acknowledge IARPA and Lincoln Labs for providing a Josephson Traveling-Wave Parametric Amplifier. We thank the SPEC at CEA Saclay for providing fabrication facilities. We thank Gerhard Kirchmair, Zaki Leghtas, Mazyar Mirrahimi, Ulysse Reglade and Marius Villiers for inspiring discussions and feedback.
\end{acknowledgments}

\appendix
\section{Circuit design}
\label{Appendix:circuit_design}

Our autoparametric circuit can be understood as a limit case of a degenerate parametric amplifier~\cite{Yurke1989, Yamamoto2008, Roy2016}. Such amplifiers are usually made of a resonator $m$ connected to a pump mode $c$ via a nonlinear element such that their interaction Hamiltonian reads $\hbar g_2 \hat{m}^2\hat{c}^\dagger+h.c.$. 

When driving the pump mode $c$ at $2\omega_m$ with a large enough power such that the number of photons in the pump mode exceeds the threshold $n_\mathrm{thr}=\kappa_m^2/(4g_2)^2$, parametric oscillation occurs. The parametric oscillation threshold corresponds to the number of photons at which the gain process compensates the losses of mode $m$. Above threshold, the number of photons in $m$ stays finite owing to one of two possible mechanisms: Kerr effect or pump depletion. The mode $m$ then emits radiation with two possible phases~\cite{Wilson2010,Wustmann2013,Svensson2018}. These two phases correspond to two coherent states $|\alpha\rangle$ and $|-\alpha\rangle$ of the mode $m$. This phenomenon can be used to stabilize cat qubits.  

The first limiting mechanism is the Kerr effect, which is a byproduct of the nonlinearity used to generate the $g_2$ rate. As the number of photons increases, the mode resonance frequency shifts up to a point where the pump at $2\omega_m$ is so detuned that the gain and loss processes balance each other. This Kerr limitation is the most standard one in Josephson circuits and is at the origin of so called Kerr cats~\cite{Puri2017, Grimm2020, Frattini2022,Iyama2023}. The second limiting mechanism is pump depletion, which is more common at optical wavelength. The stiff pump approximation breaks down if the number of photons in the pump mode is affected by how fast pump photons are consumed to generate photons in mode $m$. The rate at which pump photons are regenerated by the drive is not fast enough to produce pairs of photons in mode $m$.

The autoparametric design evades the Kerr limitation and operates in the pump depletion limit, which engineers the reservoir of mode $m$ and not its Hamiltonian. Instead of a far detuned pump mode, we use the buffer mode $b$ that resonates at the resonator frequency $2\omega_m$. The high Q limit of mode $a$ lowers the parametric oscillation threshold to $n_\mathrm{thr}\ll 1$ while the resonant driving of mode $b$ ensures that the regeneration rate of pump photons in the buffer mode is minimal. Therefore, pump depletion is the dominating mechanism for parametric oscillation stabilization. In the steady state, the buffer is subject to two opposing driving forces. The buffer drive and the action of the memory mode on the buffer via the 2-to-1 photon exchange Hamiltonian compensate exactly. The buffer then reaches a steady state close to the vacuum while the memory state converges to a superposition of $|\alpha\rangle$ and $|-\alpha\rangle$. 

This realization is at the origin of the autoparametric circuit design. We started from a superconducting circuit widely used for degenerate parametric amplification: a resonator comprising a DC flux biased SQUID that is flux pumped at twice its frequency and acts as our memory mode~\cite{Yamamoto2008}. Other non-linear elements could be chosen from the variety of Josephson amplifiers that have been designed over the past two decades. In this peculiar circuit, there exists another mode associated to the flux degree of freedom, which has a differential symmetry with respect to the SQUID junctions but often ignored for its very high frequency~\cite{Lu2023}. For our purpose, this mode, our buffer mode, is brought down in frequency while preserving its symmetry by adding an inductive element in the SQUID loop (the junction $E_W$) and splitting the capacitance on either side of the SQUID. The symmetry is preserved and used advantageously to be able to couple preferentially to the buffer mode without needing frequency selective filtering to protect the memory lifetime.  This basis structure is then diluted with open and shorted stubs as described in Fig.~\ref{fig:fig1} to tune the modes $\varphi_\mathrm{zpf}$. 

Compared to previously pumped circuits comprising an ATS~\cite{Lescanne2020exponential,Berdou2022}, this design does not ensure that the memory Kerr non-linearity vanishes at the working point, and it is hard to ensure by fabrication that $\phi_\mathrm{QEC}$ is a local extremum for the memory frequency so that flux noise affects the memory pure dephasing rate. However, on the one hand it has been shown that Kerr effect is not detrimental to cat qubit stabilization as long as it is smaller than $g_2$~\cite{Gautier2022}. This condition is here satisfied by choosing the modes $\varphi_\mathrm{zpf}$ adequately. On the other hand, cat-qubit stabilization is primarily designed to correct against dephasing~\cite{Mirrahimi2014}. The pure dephasing of the memory mode is further mitigated by design. Since the memory mode does not participate in the central inductive element of the SQUID, we implement it using a single Josephson junction acting as a weak link in the loop $E_W < E_J$ without increasing the memory self-Kerr rate. Using a single Josephson junction instead of a more linear inductance dramatically increases the buffer susceptibility to flux compared to the memory, hence the buffer is responsible for meeting the frequency matching condition $2\omega_m = \omega_b$ and the memory can afford a much weaker dependence on flux, hence on flux noise. Using a Josephson junction with a lower Josephson energy $E_W$ than the SQUID junctions $E_J$ creates sweet spots in the memory flux dispersion and the circuit parameters are designed such that the frequency matching condition occurs close to it. We provide further details on the sweet spot in the next section.

\begin{figure}[!h]
\includegraphics[width=\linewidth]{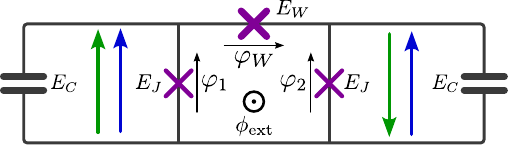}
\caption{Simplified circuit diagram of the device. Only the modes of the ring of junctions are considered for simplicity. For each junction $\varphi_x =\tilde \varphi_x + \bar \varphi_x$, respectively the oscillating part and the equilibrium part.
}
\label{fig:Res_cat_circuit}
\end{figure}

\section{Hamiltonian derivation}
\label{Appendix:Hamiltonian_derivation}

This section derives the Hamiltonian of the simplified version of the circuit, represented in Fig.~\ref{fig:Res_cat_circuit}. Since this circuit operates in the regime of small zero point fluctuations of the phase across the Josephson junctions, we will first compute the effective inductance of each junction due to the flux bias. Next, we will identify the eigenmodes of the linear part of the Hamiltonian. Finally, we will calculate the nonlinearities and discuss the impact of junction asymmetry.

\begin{figure}[!h]
\includegraphics[width=\linewidth]{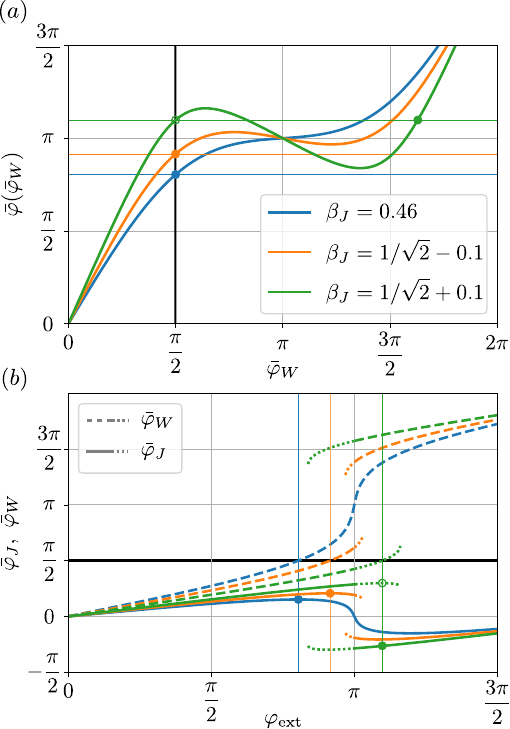}
\caption{Flux sweet spots of the mixing element. (a) Sum $\bar \varphi$ of the phase differences across the three junctions as a function of the phase difference $\bar \varphi_W$ across the weak junction, calculated using Eq.~\eqref{eq:varphisum} for $\beta_J = E_W/E_J = 0.46$ as in the experiment (blue), $\beta_J<1/\sqrt{2}$ (orange) and $\beta_J>1/\sqrt{2}$ (green). A flux bias imposes $\varphi_\mathrm{ext} = \bar\varphi$ so that these curves show which values of $\bar\varphi_W$ are possible. (b) Phase differences $\bar \varphi_J$ (solid-dotted) and $\bar\varphi_W$ (dashed-dotted) as a function of $\varphi_\mathrm{ext}$. Same color code for the values of $\beta_J$ as in (a). Dotted lines correspond to extensions of the solutions that are stable but not the lowest energy configuration. (a/b)
The black vertical/horizontal line corresponds to $\bar\varphi_W =\pi/2$. In (a) it crosses $\bar \varphi$ at $\varphi_\mathrm{ext}^\mathrm{(sweet)}$ (thin horizontal colored lines). In (b) it crosses $\bar\varphi_W$ where $\bar\varphi_J$ has a sweet spot at the same values of $\varphi_\mathrm{ext}^\mathrm{(sweet)}$ (thin vertical colored lines). The blue and orange closed circles correspond to sweet spots in the lowest energy configuration. The open green circle emphasizes that while the sweet spot exists, it is not the lowest energy configuration. At this value of $\varphi_\mathrm{ext}$ (vertical green line) another configuration is favored (green closed circle) which is not a flux sweet spot.}
\label{fig:flux_solution}
\end{figure}

\subsection{Equilibrium phase configuration}

First, we compute the effective inductance of the central mixing element. This element comprises 2 identical main junctions with Josephson energy $E_J$ and one weaker junction with energy $E_W=\beta_JE_J$ with $\beta_J<1$. Following the procedure detailed in~\cite{Miano2023}, we compute the equilibrium phase drop across each junction $\bar\varphi_1$ and $\bar \varphi_2$ and $\bar \varphi_W$.  Note that we decompose each phase drop $\varphi_x$ as a sum of a constant part $\bar\varphi_x$ and a dynamical part $\tilde\varphi_x$ with zero mean. Current conservation inside the loop imposes that $\bar \varphi_1= - \bar \varphi_2$ so that we denote $\bar \varphi_J=\bar \varphi_1$ the main junction phase drop. Current conservation further dictates that
\begin{equation}
\label{eq:varphi_J}
\bar\varphi_J = \arcsin(\beta_J\sin{\bar \varphi_W}).
\end{equation}
The phase drop around the loop $\bar \varphi = \bar \varphi_W + \bar \varphi_1 - \bar \varphi_2$ then reads
\begin{equation}
\label{eq:varphisum}
\bar \varphi = \bar \varphi_W + 2 \arcsin(\beta_J\sin{\bar \varphi_W}).
\end{equation}
Besides the superconducting loop is threaded by the external
flux $\phi_\mathrm{ext} = \varphi_\mathrm{ext} \varphi_0$, which leads to the constraint $\varphi_1 + \varphi_W - \varphi_2 = \varphi_\mathrm{ext}$. The constraint translates into
\begin{equation}
\label{eq:constraints}
\bar \varphi = \varphi_\mathrm{ext} \textrm{ and }
\tilde\varphi_1 + \tilde\varphi_W - \tilde\varphi_2 = 0 \,.
\end{equation}

Finding the configuration of phase drops at equilibrium thus consists in first determining $\bar \varphi_W(\varphi_\mathrm{ext})$ by solving the equation $\varphi_\mathrm{ext} = \bar\varphi(\bar \varphi_W)$ of which a graphical representation is shown in Fig.~\ref{fig:flux_solution}a. Then, one gets $\bar\varphi_J$ from Eq.~\eqref{eq:varphi_J} and the effective inductive energies of each junction $\bar E_J = E_J \cos(\bar\varphi_J)$, $\bar E_W = E_W \cos(\bar\varphi_W)$. The effective inductances follow as $\bar L_J = \varphi_0^2/\bar E_J$ and $\bar L_W = \varphi_0^2/\bar E_W$. In case where $\beta_J>1/2$, the function $\bar\varphi_W(\varphi_\mathrm{ext})$ is multi-valued~\cite{Miano2023} as shown in Fig.~\ref{fig:flux_solution}b. In order to avoid hysteresis effects or instabilities, we chose  $\beta_J <1/2$ which is equivalent to  $E_W<E_J/2$. We can check in Fig.~\ref{fig:flux_solution}a that $\bar \varphi(\bar \varphi_W)$ is indeed monotonous in that case.

For $\beta_J=0.46$ as in our device, at $\phi_\mathrm{QEC} = 0.311\phi_0$ that is slightly offset from the sweet spot ($0.40\phi_0$), we find $\bar \varphi_J = 0.42$ and $\bar \varphi_W = 1.11$ leading to $\bar E_J/h = 228 ~\mathrm{GHz}$ and $\bar E_W/h = 51~\mathrm{GHz}$. At $\phi_\mathrm{tomo} = 0.168\phi_0$, we find $\bar \varphi_J = 0.25$ and $\bar \varphi_W = 0.56$ leading to $\bar E_J/h = 242 ~\mathrm{GHz}$ and $\bar E_W/h = 97~\mathrm{GHz}$.  

We further study the mixing element to determine the value of the flux sweet spot to evade the memory from flux noise: It is defined as the flux $\varphi_0\varphi_\mathrm{ext}^\mathrm{(sweet)}$ such that 
\begin{equation}
\left. \frac{d\omega_m}{d\varphi_\mathrm{ext}}\right|_{\varphi_\mathrm{ext}^\mathrm{(sweet)}}= 0.
\end{equation}
We focus on the case where $\varphi_\mathrm{ext}^\mathrm{(sweet)} \in [0, \pi]$ because of the $2\pi$ periodicity of the solutions, and by symmetry around $\varphi_\mathrm{ext}=0$. Additionally, we exclude the trivial solution $\varphi_\mathrm{ext}=0$ because we need  $\bar \varphi_J \neq 0$ at the sweet spot to maintain a non-zero 3-wave mixing interaction rate. The symmetry of the circuit of Fig.~\ref{fig:Res_cat_circuit} implies that the memory mode, which is the common mode, does not participate in the weak junction. Hence the flux dependence of $\omega_m$ originates from $\bar E_J(\varphi_\mathrm{ext})=E_J \cos(\bar \varphi_J)$. The flux sweet spot is therefore a local extremum of $\bar \varphi_J(\varphi_\mathrm{ext})$. From Eq.~\eqref{eq:varphi_J}, $\varphi_\mathrm{ext}^\mathrm{(sweet)}$ is therefore close to $\bar \varphi_W = \pi/2$ since $\sin{\varphi_W}$ is maximal there. Finally, using the condition that $\varphi_\mathrm{ext} <\pi$ imposes an upper bound on $\beta_J$. Indeed, for $\bar \varphi_W = \pi/2$, Eq.~\eqref{eq:varphisum} comes down to $\pi/2 + 2\arcsin{\beta_J} < \pi$ and thus $\beta_J<1/\sqrt{2}$. 

Provided $E_W<E_J/\sqrt{2}$, a sweet spot will thus be visible at $\varphi_\mathrm{ext}^\mathrm{(sweet)} = \pi/2 + 2\arcsin\beta_J$ (blue and orange closed circles in Fig.~\ref{fig:flux_solution}b). Close to the sweet spot, at first order, $\bar \varphi_J$ is constant and $\bar \varphi_W = \pi/2 + \delta\varphi_\mathrm{ext}$  where $\delta\varphi_\mathrm{ext}=\varphi_\mathrm{ext}-\varphi_\mathrm{ext}^{(\mathrm{sweet})}$ is the flux offset from the sweet spot. At second order, we then have 
\begin{equation}
\begin{aligned}
\label{eq:sweet_spot}
\bar \varphi_J  &= \arcsin(\beta_J\sin(\pi/2 + \delta\varphi_\mathrm{ext})) \\
&= \arcsin(\frac{E_W}{E_J}(1-\frac{\delta\varphi_\mathrm{ext}^2}{2}))\,.
\end{aligned}
\end{equation}

Note that the single minimum condition $\beta_J<1/2$ that is verified in the experiment is well in this limit. The regime $1/2 < \beta_J <1/\sqrt{2}$ comprises a sweet spot in the lowest energy flux configuration (orange closed circle in Fig.~\ref{fig:flux_solution}b) but the existence of another higher energy minimum might be detrimental. Finally for $\beta_J >1/\sqrt{2}$, there exists a sweet spot when but it is not the lowest energy phase configuration (green open circle in Fig.~\ref{fig:flux_solution}b). 

\subsection{Eigenmodes}

Now that the equilibrium phase differences and the effective inductive energies $\bar E_J$ and $\bar E_W$ are determined, we can determine the eigenmodes of the system and compute the zero point fluctuation of the phase across each dipole of the linear equivalent circuit. 

The potential energy of the linear system modeling the circuit reads $U_\mathrm{lin} = \bar E_J \tilde\varphi_1^2/2 + \bar E_J \tilde\varphi_2^2/2 + \bar E_W \tilde\varphi_W^2/2$. Hence, incorporating the constraints of Eq.~\eqref{eq:constraints}, we find
\begin{equation}
\begin{aligned}
U_\mathrm{lin}\left(\tilde\varphi_1, \tilde\varphi_2 \right) &= \frac{\bar E_J}2 \tilde\varphi_1^2 + \frac{\bar E_J}2 \tilde\varphi_2^2 + \frac{\bar E_W}2 (\tilde\varphi_1 - \tilde\varphi_2)^2 \\
T\left(\dot{\tilde{\varphi}}_1, \dot{\tilde{\varphi}}_2\right) &= \frac{\hbar^2}{16 E_C}\dot{\tilde{\varphi}}_1^2 + \frac{\hbar^2}{16 E_C}\dot{\tilde{\varphi}}_2^2
\end{aligned}
\end{equation}
where T is the kinetic energy of the system and $E_C$ is the charging energy of each capacitor. This system can be readily diagonalized by performing the change of variable
\begin{equation*}
\left\{
\begin{aligned}
\varphi_m = \frac{\tilde\varphi_1 + \tilde\varphi_2}{2}\\
\varphi_b = \frac{\tilde\varphi_1 - \tilde\varphi_2}{2},
\end{aligned}
\right.
\end{equation*}
\noindent where $\varphi_m$ and $\varphi_b$ correspond to the common and differential modes of the circuit, respectively denoted as memory and buffer modes. The potential and kinetic energies of the circuit then read
\begin{equation*}
\begin{aligned}
U_\mathrm{lin}\left(\varphi_m, \varphi_b \right) &= \frac{E_{L, m}}2 \varphi_m^2 + \frac{E_{L, b}}2 \varphi_b^2\\
T\left(\dot{\varphi}_m, \dot{\varphi}_b \right) &=  \frac{\hbar^2}{8 E_C}\dot{\varphi}_m^2 + \frac{\hbar^2}{8 E_C}\dot{\varphi}_b^2.
\end{aligned}
\end{equation*}
with $E_{L, m} = 2 \bar E_J$ and $E_{L, b} = 2 \bar E_J + 4 \bar E_W$. 
The mode frequencies and zero-point fluctuations of the buffer and memory mode are given by
\begin{equation*}
\begin{matrix}
\omega_b = \sqrt{4 E_C E_{L, b}}&\varphi_{\mathrm{zpf}, b} = \left(E_C / E_{L, b} \right)^{1/4}\\
\omega_m = \sqrt{4 E_C E_{L, m}}&\varphi_{\mathrm{zpf}, m} = \left(E_C / E_{L, m} \right)^{1/4}
\end{matrix}
\end{equation*}
so that, in second quantization, the linear part of the Hamiltonian is
\begin{equation}
\begin{aligned}
\hat{H}_\mathrm{lin}/\hbar = \omega_m \hat{m}^\dagger \hat{m} + \omega_b \hat{b}^\dagger \hat{b}
\end{aligned}
\end{equation}
with annihilation operators defined by their relation to the phase differences
\begin{equation}
\begin{aligned}
\hat\varphi_m &= \varphi_{\mathrm{zpf}, m} \left(\hat{m} +  \hat{m}^\dagger \right)\\
\hat\varphi_b &= \varphi_{\mathrm{zpf}, b} \left(\hat{b} +  \hat{b}^\dagger \right).
\end{aligned}
\end{equation}

\noindent The frequencies of both the memory and buffer modes depend on $\phi_\mathrm{ext}$ through the dependency of $\bar \varphi_J$ and $\bar \varphi_W$. In the actual circuit, stub resonators are connected to the ring of junctions so that the mode frequencies and zero point fluctuations are modified. $\phi_\mathrm{ext}$ is then chosen so that the condition $\omega_b = 2 \omega_m$ is matched, defining the value of $\phi_\mathrm{QEC}$ as shown in Fig.~\ref{fig:fig1}. Numerically, fitting the frequency dispersion versus flux with the full model gives us $\varphi_{\mathrm{zpf}, m}= 0.0305 $ and $\varphi_{\mathrm{zpf}, b} = 0.0648$ at $\phi_\mathrm{QEC}$.

\subsection{Nonlinearities}

\begin{figure}[!h]
\includegraphics[width=\linewidth]{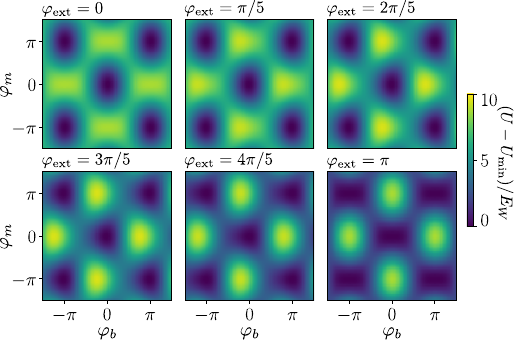}
\caption{Potential energy $U\left(\varphi_m, \varphi_b \right)/E_W$ for $E_W/h = 115~\mathrm{GHz}$ and $E_J/h = 250~\mathrm{GHz}$ at six different flux biases. For each flux bias, an offset $U_\mathrm{min}(\varphi_\mathrm{ext})$ is subtracted to Eq.~(\ref{eq:Sup_mat_potential_energy}) in order to set the potential global minimum to $0$ and better highlight the height of the potential barrier.}
\label{fig:Res_cat_potential}
\end{figure}

Around the DC solution of the system, the full potential 
\begin{equation}
\begin{aligned}
U(\tilde\varphi_1, \tilde\varphi_2) = -E_J\cos(\tilde\varphi_1 + \bar\varphi_J) -E_J\cos(\tilde\varphi_2 - \bar \varphi_J) \\ -E_W \cos(\tilde\varphi_2 - \tilde\varphi_1 + \bar\varphi_W)
\end{aligned}
\end{equation} 
can be used to compute the nonlinearities. Using the former change of variable we get
\begin{equation}
\begin{aligned}
U\left(\varphi_m, \varphi_b \right) = &-2 E_J \cos{ \left( \varphi_m\right)}\cos{ \left( \varphi_b + \bar\varphi_J \right)}  \\ 
& - E_W \cos{ \left( 2 \varphi_b - \bar \varphi_W \right)}
\label{eq:Sup_mat_potential_energy}
\end{aligned}
\end{equation}
The potential energy is represented in Fig.~\ref{fig:Res_cat_potential} for the circuit parameters $E_W/h \approx 115~\mathrm{GHz}$ and $E_J/h \approx 250~\mathrm{GHz}$ for several values of the external flux bias $\phi_\mathrm{ext}$. As expected from the change of variable, the global minimum is located at $(\varphi_m, \varphi_b) = (0,0)$ for any flux bias $\varphi_\mathrm{ext}$. Other minima exist owing to the periodicity of the Josephson potential but there is a potential barrier of $2E_J\approx k_B\times 22~\mathrm{K}$ to overcome in order to transit from one solution to the other as can be seen in Fig.~\ref{fig:Res_cat_potential}. By expanding the $\sin$ and $\cos$ functions up to fourth order in the phases, we obtain the parameters of the Hamiltonian. 

\paragraph{Third order terms}

We now consider working at $\phi_\mathrm{QEC}$ yielding the condition $\omega_b = 2 \omega_m$, and restrict our analysis to terms surviving the Rotating Wave Approximation. The only third order terms to consider is then 

\begin{equation*}
\begin{aligned}
\hat{H}_{3^\mathrm{rd}}/\hbar = E_J \sin \left(\bar\varphi_J \right) \varphi_{\mathrm{zpf}, b} \, \varphi_{\mathrm{zpf}, m}^2 \left(\hat{b} \, \hat{m}^{\dagger\, 2} + \hat{b}^\dagger \hat{m}^2\right).
\end{aligned}
\end{equation*}

\noindent This Hamiltonian corresponds to the 2-photon exchange Hamiltonian, responsible for converting single photons in the buffer into pairs of photons in the memory mode and back. Plugging in the value of $\bar\varphi_J$ around the sweet spot Eq.~\eqref{eq:sweet_spot}, we recover Eq.~\eqref{eq:g2} $\hbar g_2 \approx E_W\left(1-\delta\varphi_\mathrm{ext}^2/2\right)\varphi_{\mathrm{zpf},m}^2 \varphi_{\mathrm{zpf},b}$. From the frequency dispersion versus flux we expect $g_2/2\pi = 6.2~\mathrm{MHz}$, very close to what is experimentally extracted from Fig.~\ref{fig:Zoscillations}.

\paragraph{Fourth order terms}

Expanding even further the Hamiltonian to fourth order terms leads to the expression of the memory and buffer self-Kerr rates, as well as the cross-Kerr coupling between these 2 modes:
\begin{equation}
\label{eq:Sup_mat_4th_order_terms}
\begin{aligned}
&\hbar\chi_{m,m} = \bar E_J  \varphi_{\mathrm{zpf}, m}^4\\
&\hbar\chi_{b,b} = \bar E_J  \varphi_{\mathrm{zpf}, b}^4 + 8 \bar E_W \varphi_{\mathrm{zpf}, b}^4\\
&\hbar\chi_{m,b} = 2 \bar E_J \varphi_{\mathrm{zpf}, m}^2 \varphi_{\mathrm{zpf}, b}^2.\\
\end{aligned}
\end{equation}
Note that it is the equivalent Josephson energies $(\bar E_J, \bar E_W)$ that appears and not the bare ones $(E_J, E_W)$. 
The corresponding Hamiltonian reads 
\begin{equation*}
\begin{aligned}
\hat{H}_{4^\mathrm{th}}/\hbar = -\frac{\chi_{m,m}}{2}\hat{m}^{\dagger \, 2} \hat{m}^2 -\frac{\chi_{b,b}}{2}\hat{b}^{\dagger \, 2} \hat{b}^2 - \chi_{m,b} \, (\hat{m}^{\dagger} \hat{m})( \hat{b}^{\dagger} \hat{b}).
\end{aligned}
\end{equation*}

\noindent The effect of these spurious terms is taken into account in the simulations used in Figs.~\ref{fig:figrates} and \ref{fig:Zoscillations}. The memory self Kerr rate is directly estimated at $\phi_\mathrm{tomo}$, $\chi_{m,m} / 2\pi \approx 220 \, \mathrm{kHz}$ from the dynamics of a coherent state. The parameters $\chi_{b,b}$ and $\chi_{m,b}$ are not measured in this experiment and instead inferred from the other measured variables using the expressions above. At $\phi_\mathrm{QEC}$, from the frequency dispersion versus flux, we expect $\chi_{m,m}/2\pi = 200~\mathrm{kHz}$, $\chi_{b,b}/2\pi = 11.3~\mathrm{MHz}$ and $\chi_{m,b} = 1.8~\mathrm{MHz}$.

\begin{figure}[h]
    \centering
    \includegraphics[width=64mm]{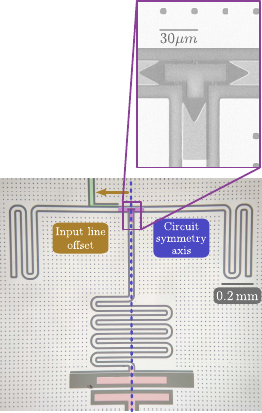}
    \caption{Optical image of the autoparametric circuit showing its symmetry axis in blue, and the offset of the input line (green) with respect to the symmetry axis in orange. Inset: larger view of the coupler and its connections to the stubs.}
    \label{fig:symmetry_axis}
\end{figure}

\begin{figure*}[!t]
    \includegraphics[width=\linewidth]{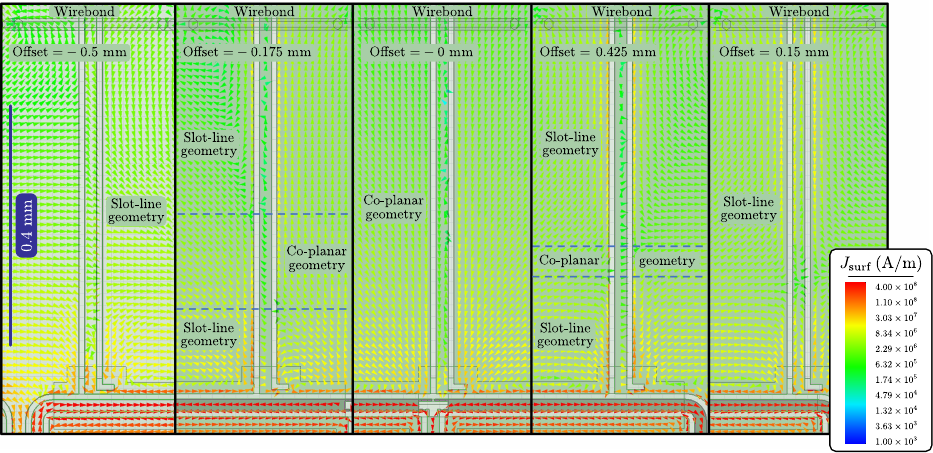}
    \caption{Electromagnetic simulations (using Ansys HFSS) of the current field of the memory mode around the input line for various offsets of the input line with respect to the circuit symmetry axis. Colors indicate surface currents according to the legend. For large offsets (-500 $\mathrm{\mu m}$ and 500 $\mathrm{\mu m}$), the current field is characteristic of a slot-line geometry (opposite current in the two ground planes) indicating that the memory is mostly coupled to the slot-line propagating mode. For zero offset,  the current field is characteristic of a co-planar geometry (identical current on the two ground planes and opposite current in the central track) indicating that the memory is mostly coupled to the co-planar waveguide propagating mode. For the two optimal offsets (-175 $\mathrm{\mu m}$ and 425 $\mathrm{\mu m}$) leading to the highest memory coupling quality factor, the current field is a mix of slot-line and co-planar geometries.}
    \label{fig:current_field}
\end{figure*}

\section{Microwave Design}
\label{Appendix:Circuit_design}

\subsection{Input line}

Because the memory mode (respectively buffer mode) is symmetric (respectively anti-symmetric, see Fig.~\ref{fig:fig1}c) with respect to the circuit symmetry axis (blue in Fig.~\ref{fig:symmetry_axis}), one can use the properties of transmission line modes of the input to disable the coupling to the memory mode. Memory and buffer modes only couple to propagating modes having the same symmetries.

A first input line design that can benefit from these symmetries is a slotline made of a gap, separating two ground planes~\cite{Pozar}. This transmission line has a single propagating mode, which is anti-symmetric since opposed currents flow in the two ground planes. When the slotline is aligned with the circuit symmetry axis, only the buffer mode is coupled to the transmission line while the memory mode is protected by symmetry. If this input line design is optimal from a point of view of memory filtering, it is not compatible with fast flux bias as it cannot bring a DC current close to the Josephson junction loop.

In contrast, a co-planar waveguide (CPW) transmission line contains a center track so that it can be used for fast flux biasing the loop. A CPW transmission line contains two quasi-transverse electromagnetic propagating modes \cite{simons2004coplanar}. One is symmetric with respect to the transmission line axis with identical currents in the ground planes but an opposite one in the central line. We call it the co-planar waveguide mode. The other is anti-symmetric with respect of the transmission line axis with opposing currents in the two ground planes and no current in the central line. We call it the slot line mode. Importantly, the slot line mode is suppressed by increasing the density of wirebonds that connect the two ground planes.

When the CPW transmission line is aligned on the circuit symmetry axis, the buffer (respectively memory) mode is only coupled to the slot-line mode (respectively co-planar waveguide mode) by symmetry. 
The mutual inductance of 2 pH that we design for flux biasing sets the geometry of the input line close to the loop. However, the position and density of wirebonds can be chosen. In practice, we tune the buffer coupling rate using wirebond positions as they affect the slot-line mode. In contrast, lowering the memory coupling rate requires another control parameter.

To this aim, we offset the CPW transmission line away from the circuit symmetry axis (orange arrow in Fig.~\ref{fig:symmetry_axis}). The buffer mode stays dominantly coupled to the slot-line mode whereas the memory mode is coupled to both slot-line and co-planar waveguide modes. This can be seen in electromagnetic simulations of Fig.~\ref{fig:current_field} by looking at the currents of the memory mode around the input line. The memory mode coupling to the co-planar waveguide (respectively slot-line) mode decreases (respectively increases) with the shift length. This can be observed by looking at the dominant geometry of the memory currents around the input line (see Fig.~\ref{fig:current_field}).

\begin{figure}
    \centering
    \includegraphics[width=86mm]{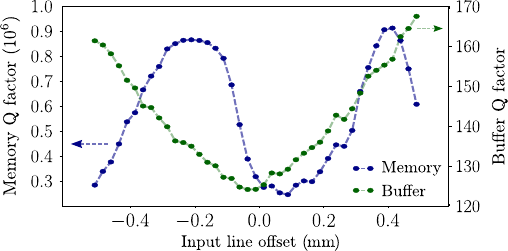}
    \caption{Simulated memory (blue) and buffer (green) coupling quality factors (Q) as a function of the input line offset with respect to the circuit symmetry axis. The memory Q factor reaches maximums for two sweet spots defined as offset lengths of 425 $\mathrm{\mu m}$ and -175 $\mathrm{\mu m}$.}
    \label{fig:Q_vs_shift}
\end{figure}

The memory coupling rate to the transmission line is given by the sum of the coupling rates to the co-planar waveguide and slot-line modes and can be simulated. Fig.~\ref{fig:Q_vs_shift} shows the variation of the simulated memory coupling quality factor $Q$ as a function of the input line offset. There are two optimal offsets (a positive one $l_+ = 425 ~\mathrm{\mu m}$ and a negative one $l_- = -175 ~\mathrm{\mu m}$) for which the total coupling rate to the slot-line and co-planar waveguide modes is minimized leading to an enhancement of the Q factor of the memory.  The $l_-$ offset has a large enough mutual inductance (about 2 pH) for fast flux biasing and it is the one we chose.

\subsection{Coupling between memory and transmon}
\label{sec:ind_coupling}

\begin{figure}[h]
    \centering
    \includegraphics[width=\columnwidth]{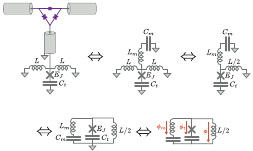}
    \caption{Equilavent electronic schemes for the inductive coupling between the transmon mode and the memory mode.}
    \label{fig:inductive_coupling}
\end{figure}

The bottom stub of the memory mode in Fig.~\ref{fig:symmetry_axis} is terminated to the ground by three elements in parallel: two identical geometrical inductances $L$ (horizontal lines) and a Josephson junction of Josephson energy $E_J$ in series with a capacitor $C_t$ (large pad in pink). The important parts of this circuit are schematized in Fig.~\ref{fig:inductive_coupling}. Focusing on the sole memory mode for the autoparametric circuit, one models it by an inductor $L_m$ and a capacitor $C_m$ in series. With the last scheme of Fig.~\ref{fig:inductive_coupling}, one sees how the small inductance $L/2$ inductively couples the memory and transmon modes. A simple way to compute this coupling term consists in writing Kirchoff's law
\begin{equation}
    \frac{\phi}{L/2}=-\frac{\phi_m}{L_m}-\frac{E_J}{\varphi_0}\sin(\phi_t/\varphi_0).
\end{equation}
We see that the inductive energy of the small inductance $L/2$ reads
\begin{equation}
    U_L= \frac{\phi^2}{L}=\frac{\left(\frac{L}{2L_m}\phi_m+\frac{L E_J}{2\varphi_0}\sin(\phi_t/\varphi_0)\right)^2}{L}.
\end{equation}
Expanding this square term leads to a renormalization of the frequencies of the memory mode and of the transmon mode and the cross-product gives the coupling term we wanted to compute:
\begin{equation}
    \hat{H}_\mathrm{ind}= \frac{L E_J}{2L_m}\frac{\hat{\phi}_m}{\varphi_0}\sin(\hat{\phi}_t/\varphi_0)= \hbar g_{mt} \sin{\hat{\theta}_t}(\hat{m}+\hat{m}^\dagger),
\end{equation}
where $\hat{\theta_t}=\hat{\phi_t}/\varphi_0$ and  $g_{mt}$ is given by $\frac{L E_J}{2\hbar L_m\varphi_0}\varphi_\mathrm{ZPF,m}$ up to a factor of order 1. The coupling rate between the memory and transmon modes, $g_{mt}$, was measured from the system's low-energy spectrum to be $2\pi\times 225~\mathrm{MHz}$. This value agrees well with the coupling rate predicted by electromagnetic simulations.

\section{Cabling}
\label{Appendix:Cabling}

\begin{figure*}[!h]
\includegraphics[width=\linewidth]{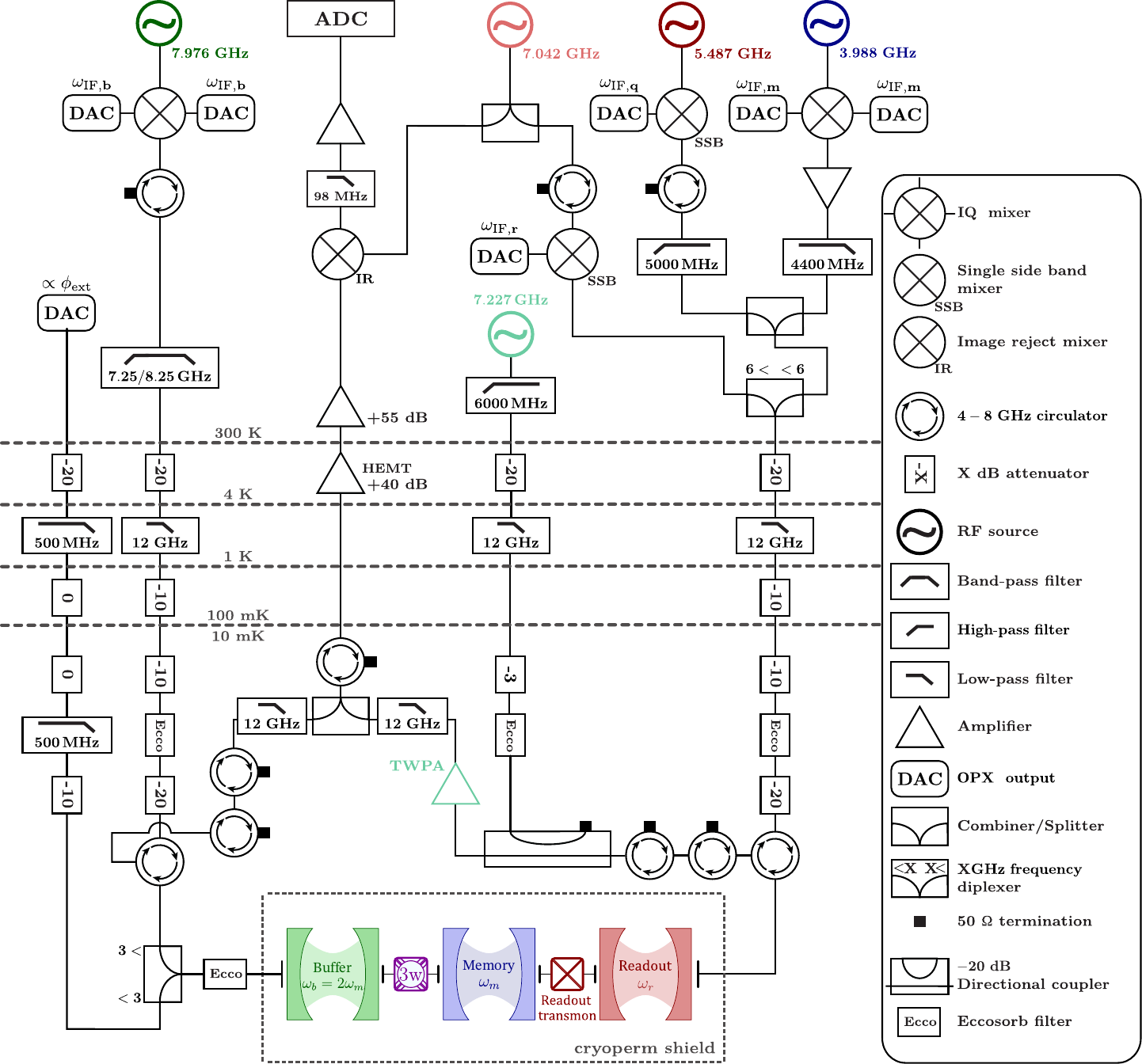}
\caption{Schematic of the setup. Each electromagnetic mode in the circuit is driven by an RF source detuned by the modulation frequency and whose color matches that of the corresponding mode.}
\label{fig:Cablage}
\end{figure*}

The transmon qubit, readout resonator, memory, and buffer modes are driven by pulses whose envelope is generated using an Arbitrary Waveform Generator (AWG), an OPX by Quantum Machine in this experiment. These pulses are respectively modulated at  $\omega_{\mathrm{IF}, q}/2\pi = 100 \, \text{MHz}$, $\omega_{\mathrm{IF}, r}/2\pi = 75 \, \text{MHz}$, $\omega_{\mathrm{IF}, m, \mathrm{tomo}}/2\pi = 40 \, \text{MHz}$ or $\omega_{\mathrm{IF}, m, \mathrm{QEC}} = \left(\omega_{m, \mathrm{tomo}}  + \omega_{\mathrm{IF}, m, \mathrm{tomo}} \right) - \, \omega_{b, \mathrm{QEC}}/2$, and $\omega_{\mathrm{IF}, b, \mathrm{QEC}} = 2\omega_{\mathrm{IF}, m, \mathrm{QEC}}$. $\omega_{\mathrm{IF}, m, \mathrm{tomo}}$ and $\omega_{\mathrm{IF}, m, \mathrm{QEC}}$ are the modulation frequencies used to respectively drive the memory at $\phi_\mathrm{tomo}$ or $\phi_\mathrm{QEC}$, $\omega_{m, \mathrm{tomo}}$ and $\omega_{b, \mathrm{QEC}}$ are the frequencies of the memory mode at $\phi_\mathrm{tomo}$  and the buffer mode at $\phi_\mathrm{QEC}$. The above condition on $\omega_{\mathrm{IF}, m, \mathrm{QEC}}$ and $\omega_{\mathrm{IF}, b}$ ensures the phase stability of the encoded cat in the frame rotating at the memory frequency.

These signals are up-converted using single sideband mixers for the transmon qubit and readout resonator, and IQ mixers for the memory and buffer mode, with Radio Frequency signals generated by a 4-channel Anapico APUASYN20. The signals at frequencies $\omega_q$, $\omega_r$, and $\omega_{m, \mathrm{tomo}}/\omega_{m, \mathrm{QEC}}$ are all combined and then sent via the readout port of the device using a 6~GHz frequency diplexer. The memory drive subsequently traverses the transmon qubit, readout resonator, and its  Purcell filter before reaching the memory cavity. Given this complex path, employing a room-temperature amplifier becomes indispensable for achieving displacements $\hat{D}(\beta)$ where $\beta \geq 2$.

The signal driving the buffer mode at $\omega_b$ is transmitted through the alternative port of the device. It's combined with a DC signal directly generated by one DAC of the OPX using a 3~GHz frequency diplexer, facilitating a swift transition from $\phi_\mathrm{tomo}$ to $\phi_\mathrm{QEC}$. We attempted to drive the memory through this port to bypass the previously described elements, but the protection from the symmetry of the non-linear coupler (Fig.~\ref{fig:fig1}c) is excessively effective and prevented the achievement of large enough displacements.

The two reflected signals from the buffer and readout modes merge at the mixing chamber. The latter is first pre-amplified by a Travelling Wave Parametric Amplifier (TWPA)~\cite{Macklin2015}. Further amplification is performed by a High Electron Mobility Transistor at the 4K stage, and then a room-temperature amplifier. Subsequently, the signal is down-converted using an image reject mixer, followed by filtering, amplification, and acquisition by an ADC of the OPX. With its capacity for real-time digitization and demodulation, the OPX allows for real-time feedback and implementation of the transmon qubit reset at the beginning of each pulse sequence. The complete setup is depicted in (Fig.~\ref{fig:Cablage}).

\section{Wigner measurement of cat states}
\subsection{The use of a fast flux line}

As explained in the main text, the preparation of the cat state $|C_+^\alpha\rangle\propto|\alpha\rangle+|-\alpha\rangle$ is as simple as starting from the memory vacuum state at $\phi_\mathrm{QEC}$ and turning on a drive with the right amplitude $|\epsilon_d|=\alpha^2 g_2$ at twice the memory frequency $\omega_d=2\omega_{m, \mathrm{QEC}}$. This drive, resonant with the buffer mode, injects photons with energy $\hbar \omega_{b, \mathrm{QEC}}$ which are converted into pairs of photons in the memory. By adiabatically eliminating the buffer, we obtain the desired effective memory dynamics, characterized by the loss operator
\begin{equation*}\hat{L}_2 = \sqrt{\kappa_2}\left( \hat{m}^2 - \alpha^2 \right),
\end{equation*}
with $\kappa_2 = 4 g_2^2/\kappa_b$. 

Our task is to measure the Wigner function $W$ of the encoded state. The conventional technique for measuring $W\left(\beta \right)$~\cite{Davidovich1996,Bertet2002,Vlastakis2013} begins by displacing the memory by $\hat{D}\left(-\beta \right)$ before applying an unconditional $\pi/2$ pulse on the transmon qubit. The system then remains idle for a time period of $\pi / \chi \approx 2.8 \mu \mathrm{s}$, during which the qubit gains information about the parity of the number of photons in the memory. Subsequently, a second $\pi/2$ pulse is applied to map the memory parity into the states $\ket{g}$ or $\ket{e}$ of the qubit.

However, this method would not be effective at $\phi_\mathrm{QEC}$ due to the large 2-photon dissipation. This would impede proper displacements $\hat{D}\left(-\beta \right)$. Moreover, this dissipation broadens the memory energy levels by $\kappa_2 \gg \chi$, effectively neutralizing the dispersive coupling between the memory and transmon qubit. 

To overcome these challenges, the Wigner tomography is performed at $\phi_\mathrm{tomo}$ where $\omega_{b, \mathrm{tomo}} \neq 2\omega_{m, \mathrm{tomo}}$. At this flux, two-photon dissipation is inactive because the 2-photon exchange Hamiltonian is not preserved in the rotating wave approximation, enabling the usual Wigner tomography. To rapidly switch between $\phi_\mathrm{QEC}$ and $\phi_\mathrm{tomo}$, we employ a fast flux line that sets the desired magnetic flux in approximately $20~\mathrm{ns}$.

While the memory dynamics at $\phi_\mathrm{tomo}$ are primarily dominated by the self-Kerr rate $\chi_{m,m}/2\pi \approx 220 \mathrm{kHz}$, which only marginally impacts the system during the 20~ns it takes to switch the flux, it is crucial to keep the drive $\epsilon_d\left(\alpha \right)$ on before shifting from $\phi_\mathrm{QEC}$ to $\phi_\mathrm{tomo}$. The memory dynamics at $\phi_\mathrm{QEC}$ are indeed dominated by the 2-photon dissipation with a rate $\kappa_2$, which significantly impacts the system in 20~ns. To prevent state distortion prior to the Wigner tomography, the drive $\epsilon_d\left(\alpha \right)$ is thus maintained during the flux change. This drive at $\omega_{b, \mathrm{QEC}}$ does not affect the memory at $\phi_\mathrm{tomo}$, where the frequency matching condition is no longer satisfied.

\subsection{Phase correction of the stabilized cat}
\label{Appendix:Cat_Wigner_phase_correction}

\begin{figure}[!ht]
\begin{center}
\includegraphics[width=0.9\linewidth]{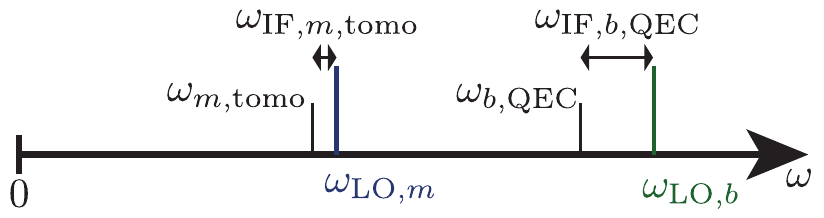}
\caption{Repartition of the frequencies of local oscillators generated by the APUASYN20 synthetizer, intermediate frequencies generated by the OPX DACs, and resonance frequencies of the device.}
\label{fig:freqs}
\end{center}
\end{figure}

Owing to the change in memory frequency when the flux is switched between $\phi_\mathrm{QEC}$ and $\phi_\mathrm{tomo}$, a carefully designed driving sequence must be followed in order to track the reference frame of the cat qubit. We set a local oscillator at $\omega_{\mathrm{LO},m} = 2\pi\times 3.988481~\mathrm{GHz}$ and another one at twice this frequency $\omega_{\mathrm{LO}, b} = 2\omega_{\mathrm{LO}, m}$ (see Fig.~\ref{fig:freqs}). They are generated using 2 channels of an Anapico APUASYN20 so that their phases are locked.

The memory displacement pulse applied for Wigner tomography at the flux $\phi_\mathrm{tomo}$ is generated by mixing the local oscillator at $\omega_{\mathrm{LO}, m}$ with a pulse generated by the AWG at exactly $\omega_{\mathrm{IF},m,\mathrm{tomo}}/2\pi=40~\mathrm{MHz}$. In contrast,
at the flux $\phi_\mathrm{QEC}$, the cat qubit is stabilized with a buffer drive at a frequency $\omega_{b, \mathrm{QEC}}=2\omega_{m, \mathrm{QEC}}$ that is not given by $2(\omega_{\mathrm{LO}, m} - \omega_{\mathrm{IF},m,\mathrm{tomo}})$ owing to the frequency change of the memory between the two flux working points. The drive at $\omega_{m, \mathrm{QEC}}$ is in fact generated by mixing the local oscillator at $2\omega_{\mathrm{LO}, m}$ with a pulse generated by the AWG at $\omega_{\mathrm{IF}, b, \mathrm{QEC}} = 2(\omega_{\mathrm{LO}, m} - \omega_{m, \mathrm{QEC}})\approx 2\pi\times 120~\mathrm{MHz}$.

\begin{figure}[!ht]
\begin{center}
\includegraphics[width=\linewidth]{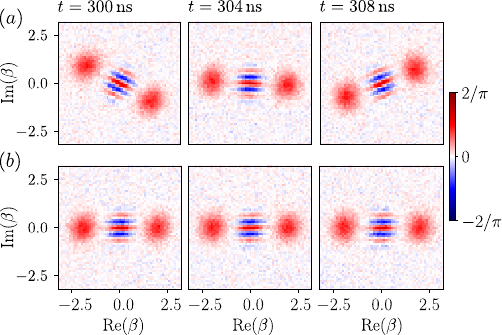}
\caption{Wigner Tomography of the cat state $\ket{C_\alpha^+}$ with $\alpha=2$ for 3 stabilization times: 300, 304 and 308~ns. (a) Evolution of the cat state when no compensation is applied. (b) Evolution of the cat when the accumulated phase between the 2 local oscillators is taken into account before memory displacement.}
\label{fig:Unstable_cat}
\end{center}
\end{figure}

Owing to the detuning of about 20~MHz between $\omega_{\mathrm{IF}, b, \mathrm{QEC}}/2$ and $\omega_{\mathrm{IF},m,\mathrm{tomo}}$, the stabilized coherent states $|\pm\alpha\rangle$ of the cat qubit are offset by a time increasing phase in the frame of the displacement pulses used for Wigner tomography. When trying to perform the desired displacement $\hat{D}(-\beta) = \hat{D}(-|\beta|e^{i\theta})$ for the Wigner tomography, this accumulated phase induces a displacement with an angle 
\begin{equation*}
\theta_\mathrm{offset}(t) = \theta +  \left(\omega_{\mathrm{IF}, b, \mathrm{QEC}}/2 - \omega_{\mathrm{IF}, m, \mathrm{tomo}} \right)t,
\end{equation*}
\noindent where $t$ is the time spent at $\phi_\mathrm{QEC}$. This can be seen as cat states whose direction in phase-space changes over time (Fig.~\ref{fig:Unstable_cat}a).  

Taking this phase offset into account, we compensate the accumulated phase directly on the AWG to keep the orientation of the cat qubit states constant in phase space when reconstructing its Wigner Tomography (Fig.~\ref{fig:Unstable_cat}b). This is of particular interest for the measurement of $T_\mathrm{X}$ where we need to measure the evolution of $W\left(\pm \alpha \right)$, which can then be done by measuring only 2 points of the Wigner function, greatly speeding up this already time-consuming measurement.

\section{Different methods to calibrate memory displacements}

\begin{figure}[!h]
\includegraphics[width=\linewidth]{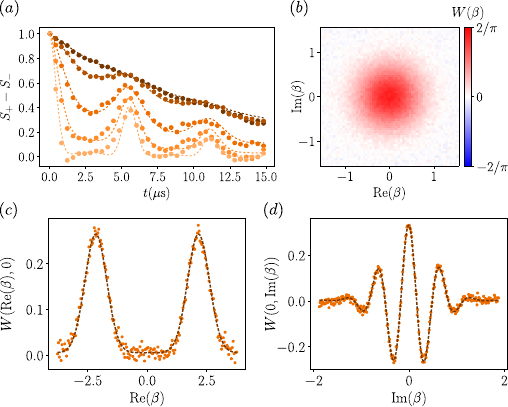}
\caption{(a) Ramsey interferometry. Dots: measured signal $S_+ - S_-$ between 2 Ramsey-like experiments for various voltages $V_d=0$, $10$, $20$, $30$, $40,~\mathrm{mV}$ from top to bottom. Lines: Fit of the measurements to Eq.~(\ref{eq:popRamsey}) leading to a photon number $\bar{n}=0.01$, $0.10$, $0.35$, $0.81$, $1.45$. The residual thermal population is thus $n_\mathrm{th}=0.01$. (b) Measured Wigner function of the memory in thermal equilibrium with its environment. The conversion used between $V_d$ and $|\beta|$ to plot it is made by a Gaussian fit of the measurement with the Wigner function of a thermal state with $n_\mathrm{th} = 0.01$ photons on average. (c) Dots: Cuts of the Wigner tomography of a stabilized cat qubit along $\beta\in\mathbb{R}$ after $100~\mu\mathrm{s}$ of dephasing. Line: Theoretical prediction. (d) Same plot along $\beta\in i\mathbb{R}$, $500~\mathrm{ns}$ after the buffer drive is turned on.}
\label{fig:Disp_calibration}
\end{figure}

The displacements $\hat{D}(\beta)$ applied on the memory during Wigner tomography are performed by applying a drive at frequency $\omega_\mathrm{m, \, tomo}$. We calibrate how the displacement amplitudes $\beta$ depend on the voltage amplitude $V_d$ at the level of the DAC by 3 methods (Fig.~\ref{fig:Disp_calibration}). We then verify how good is the match between the proportionality factor $\mu=\mathrm{d}\beta/\mathrm{d}V_d$ they provide. 

\subsection{Ramsey interferometry}

The first calibration method relies on a Ramsey sequence~\cite{Campagne2015}. Starting from the memory in its vacuum state, a drive of amplitude $V_d$ is applied to displace the memory to a coherent state $\ket{\beta}$. Accounting for the residual thermal occupation of the memory mode, the mean number of photons is $\bar{n}= \beta^2+n_\mathrm{th}$. The dispersively coupled qubit is then prepared in an equal superposition of ground and excited state by applying an unconditional $\pi/2$ pulse. After a varying time $t$, the superposition accumulates a phase $\chi \hat{m}^\dagger\hat{m}t$ that depends on the memory photon number $\hat{m}^\dagger\hat{m}$. A second unconditional $\pm \pi/2$ pulse is then applied on the qubit, which is then measured to give 2 average signals $S_{\pm}$. The difference between these 2 signals then evolves as~\cite{Campagne2015} 
\begin{equation}
    S_+ - S_- = \cos\left(\bar{n} \sin\left(\chi t \right)\right) e^{\bar{n} \left( \cos\left( \chi t\right) -1\right) -t/T_2}.\label{eq:popRamsey}
\end{equation}

\noindent From this measurement (Fig.~\ref{fig:Disp_calibration}a), we can extract the cross-Kerr coupling rate $\chi/2\pi = 170 \mathrm{kHz}$ between memory and transmon qubit. We also obtain the thermal population $n_\mathrm{th} = 0.011\pm0.002$ and $\mu = 31.33 \pm 0.85~\mathrm{V}^{-1}$.

\subsection{Thermal state tomography}
\label{sec:thermal_population}

Another calibration method is to perform a Wigner tomography of the memory thermal state, using the independently measured average occupation $n_\mathrm{th} = 0.01$. The density matrix can be written as a Boltzmann distribution $\hat{\rho}_\mathrm{th} = \sum_n \frac{n_\mathrm{th}^n}{\left(1+n_\mathrm{th} \right)^{n+1}} \ket{n}\bra{n}$ and the Wigner function associated 

\begin{equation}
    W_\mathrm{th}\left( \beta \right) = \sum_n \frac{n_\mathrm{th}^n}{\left(1+n_\mathrm{th} \right)^{n+1}} W_n \left( \beta \right),
\end{equation}

\noindent where $W_n \left( \beta \right) = \left(-1\right)^n\frac{2}{\pi}e^{- 2\left|\beta\right|^2} L_n \left(4 \left|\beta\right|^2 \right)$ is the Wigner function of Fock state $\ket{n}$. Laguerre polynomials obey the following rule

\begin{equation*}
    \sum_n t^n L_n \left(4 \left|\beta\right|^2 \right) = \frac{1}{1-t}e^{4t\left|\beta\right|^2 / (1-t)}.
\end{equation*}

\noindent Therefore, we obtain 

\begin{equation}
\begin{aligned}
    W_\mathrm{th}\left( \beta \right) &= \frac{2}{\pi \left(1+n_\mathrm{th} \right)}e^{-2\left|\beta\right|^2} \sum_n \left( \frac{ -n_\mathrm{th}}{1+n_\mathrm{th}} \right)^n L_n \left(4\left|\beta\right|^2 \right)\\
    &= \frac{2}{\pi \left(1+2n_\mathrm{th} \right)}e^{-2\left|\beta\right|^2/\left(1+2n_\mathrm{th} \right)}.
\end{aligned}
\end{equation}

\noindent Using a conversion factor $\mu = 31.31 \pm 0.14~\mathrm{V}^{-1}$ rescales the displacement amplitudes from voltages $V_d$ into complex amplitudes $\beta$ for the measured Wigner function in (Fig.~\ref{fig:Disp_calibration}b), in such a way that the standard deviation $\sigma$, of this Gaussian distribution is $\sigma = \sqrt{1 + 2n_\mathrm{th}}/2$, with $n_\mathrm{th} = 0.01$.

\subsection{Measurement of cat states fringes}

Our last method to calibrate the conversion factor is based on the Wigner tomography of a cat state~\cite{Milul2023}. The particular features of the cat Wigner function allow to directly estimate $\mu$, assuming the distortion due to memory self-Kerr or thermal population is negligible. The Wigner function of an even cat state of size $\alpha$, $\ket{C_\alpha^+}$, can be written as 

\begin{equation}
\begin{aligned}
    W_\alpha^+\left(\beta \right) = \frac{1}{\pi} &\left(  e^{-2 \left|\alpha - \beta \right|^2} +  e^{-2 \left|\alpha + \beta \right|^2} \right. \\ 
    & + \left. 2 \cos\left( 4 \mathrm{Im}\left(\alpha^* \beta \right)\right)e^{-2 \left|\beta \right|^2} \right).
\end{aligned}
\end{equation}

\noindent Introducing $\delta V_\alpha$ and $\delta V_\mathrm{I}$ the drive voltages corresponding to respectively a displacement of $2\alpha$ (distance between the two Gaussian distributions in Fig.~\ref{fig:Disp_calibration}c) and the periodicity of the fringes $\pi / 2\alpha$ (seen in Fig.~\ref{fig:Disp_calibration}d), this yields 

\begin{equation*}
\left\{
\begin{aligned}
    2 \alpha &= \mu \, \delta V_\alpha \\
    \frac{\pi}{2 \alpha} &= \mu \, \delta V_\mathrm{I}\\
\end{aligned}
\right.
\end{equation*}

Therefore $\mu = \sqrt{\pi / \left(\delta V_\alpha  \delta V_\mathrm{I}\right)}$. The values of $\delta V_\alpha$ and $\delta V_\mathrm{I}$ are measured via cuts of the Wigner function in the direction and orthogonal to the cat state  (Fig.~\ref{fig:Disp_calibration}c,d). The conversion factor obtained via this method is $\mu = 31.41 \pm 0.04~\mathrm{V}^{-1}$, once again compatible with the previous calibrations. With the conversion factor being calibrated, this method can actually be used in order to estimate the cat size by simply looking at the fringes' periodicity, as has been done in~\cite{Milul2023}.

\subsection{Calibration}

To conclude this section, the three methods are compatible with $\mu=31.4\pm 0.1~\mathrm{V}^{-1}$. We use this value of $\mu=31.4~\mathrm{V}^{-1}$ for the whole article.

\section{Measurement of $\kappa_2$ and $\kappa_1$}
\label{Appendix:fit_kappas}

\subsection{Determination of $\kappa_2$ using engineered relaxation of cat qubits}

In order to measure the rate $\kappa_2$, we first prepare $\ket{C_\alpha^+}$ or $\ket{C_\alpha^-}$ by driving the buffer with a drive $\epsilon_d \left( \alpha \right)$ at $\phi_\mathrm{QEC}$. Turning off $\epsilon_d \left( \alpha \right)$ while remaining at $\phi_\mathrm{QEC}$ then ensures the memory loses pairs of photons to the environment at the rate $\kappa_2$. $\ket{C_\alpha^\pm}$ then converges to a state in the manifold $\{ \ket{0}, \ket{1} \}$ with the same parity as the initial state. An example of such an evolution starting from $\ket{C_\alpha^+}$, $\alpha = 2.5$, is shown at a few decay times in Fig.~\ref{fig:fig2}. The complete list of measured decay times for this evolution is 
t = 0, 4, 8, 12, 20, 28, 40, 60, 100, 160, 240 and 320~ns.

In order to extract the rate $\kappa_2$ from these dynamics, the initial density matrix describing the memory is approximated by

\begin{equation}
    \rho = p\ket{C_{\alpha}^+}\bra{C_{\alpha}^+} + (1-p) \ket{C_{\alpha}^-}\bra{C_{\alpha}^-}.\label{eq:mixedcat}
\end{equation}

\noindent $\alpha$ is extracted by fitting the initial measured Wigner tomography, while $p$ is deduced from the value of $W(0)$ which fully characterizes the parity of the state. However, the obtained description of the initial density matrix is only an approximation as it does not take into account possible leakage out of the code space due to the memory self Kerr effect, dispersive coupling to transmon and buffer modes, or potential heating effect. From this initial state (\ref{eq:mixedcat}), the evolution of the memory state is then simulated using the Hamiltonian and loss operators 

\begin{equation}
\begin{aligned}
    &\hat{H} / \hbar = -\frac{\chi_{m,m}}{2}\hat{m}^{\dagger 2} \hat{m}^2 \\
    &\hat{L}_1 = \sqrt{\kappa_1}\hat{m}, \ \hat{L}_2 = \sqrt{\kappa_2}\hat{m}^2.
\end{aligned}
\end{equation}

\noindent The single photon loss rate $\kappa_1/2\pi \sim 14~\mathrm{kHz}$ is extracted from the decay of the single photon state $\ket{1} \rightarrow \ket{0}$ (see section~\ref{kappa1determination}). Using the memory self-Kerr rate of $2\pi\times 220~\mathrm{kHz}$ measured at $\phi_\mathrm{tomo}$, we use its predicted flux dependence in Eq.~(\ref{eq:Sup_mat_4th_order_terms}) to estimate that the self-Kerr rate at $\phi_\mathrm{QEC}$ is $\chi_{m,m}/2\pi \sim 206~\mathrm{kHz}$. Minimizing the difference between measured and simulated Wigner functions at all times $t$,
\begin{equation}
    \sum_\mathrm{t} \ \int_\mathbb{C} \, \left| W_\mathrm{exp}\left( \beta,t \right) - W_\mathrm{sim}\left( \beta,t \right)\right| \mathrm{d}\beta,
\end{equation}
then allows us to fit the value of $\kappa_2$ that best reproduces the memory dynamics. 

\begin{figure}[!h]
\includegraphics[width=\linewidth]{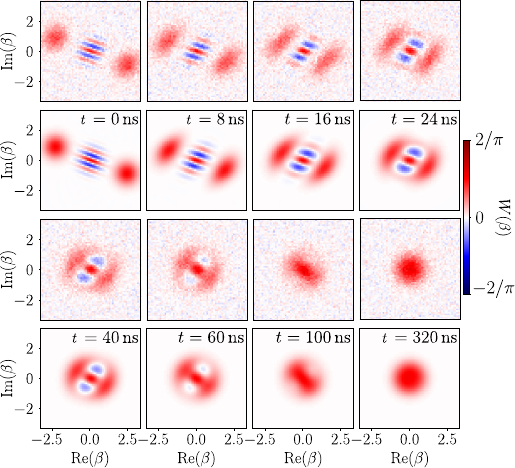}
\caption{Evolution of the Wigner functions of the memory starting close to a cat state $\ket{C_\alpha^+}$, under the effect of two-photon dissipation at $\phi_\mathrm{QEC}$, without driving the buffer. First and third lines: measured Wigner functions at various times indicated on the figure. Second and fourth lines: simulated Wigner functions of the memory without the adiabatic elimination of the buffer, and with a two-photon coupling rate $g_2/2\pi = 6~\mathrm{MHz}$ (Fig.~\ref{fig:figrates}e)}
\label{fig:Simu_vs_data_deflate_non_adiab}
\end{figure}

The uncertainty shown in Fig.~\ref{fig:fig2}a is then calculated using the result of the minimization method $\Delta \kappa_2 \leq \sqrt{\mathrm{tol} \, \times \, H^{-1}}$, with tol being the tolerance given for the convergence of the algorithm and $H^{-1}$ the inverse of the Hessian matrix. It should be noted that, due to the condition for the adiabatic elimination of the buffer 

\begin{equation}
    8 g_2 \alpha < \kappa_b
\end{equation}

\noindent not being verified for $\alpha \gtrsim 1$, we observe a deviation between the experimental data in Fig.~\ref{fig:fig2}b and the evolution predicted by this simple model. Actually, the buffer mode gets populated during two-photon dissipation. In turn, the memory sees an effective drive originating from this buffer population, inducing small deformations of the Wigner function. Particularly visible at 8~ns where the buffer is close to being maximally populated, this effect vanishes at 40~ns after the memory loses enough photons for the system to re-enter the adiabatic regime. This effect can be taken into account in the simulation by including the buffer dynamics without adiabatic elimination (Fig.~\ref{fig:Simu_vs_data_deflate_non_adiab}). In practice, we use the same model as in Sec.~\ref{sec:kappaztheo} apart from the detunings $\Delta_m=1~\mathrm{MHz}$ and $\Delta_b=0$. However this bipartite evolution does not provide an effective value of $\kappa_2$ acting on the memory mode, hence the benefit to stick with the simpler model. 

We have tried other methods to estimate $\kappa_2$, in particular by extracting $\left< \hat{m}^\dagger\hat{m}\right>$ from the measured Wigner function 

\begin{equation}
\Bar{n} =  \int_\mathbb{C} W_\mathrm{exp}\left( \beta \right) \beta^2 \mathrm{d}\beta - 0.5
\end{equation}

\noindent and compare it with the theoretical expression given in~\cite{H_D_Simaan_1975}. However, reconstructing $\Bar{n}$ with this method has proven quite challenging due to the measurement noise of the Wigner tomography, which would have made it necessary to use Maximum Likelihood Estimation (MLE)~\cite{Sivak2022} in order to circumvent this issue. 

\subsection{Determination of $\kappa_1$ using the relaxation of a single photon}

\label{kappa1determination}
\begin{figure}[!h]
\includegraphics[width=\linewidth]{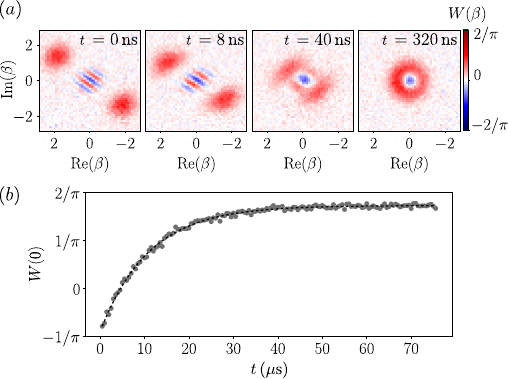}
\caption{(a) Measured Wigner functions of the memory starting close to $\ket{C_\alpha^-}$ after the decay times indicated on the figure. (b) Dots: Measured evolution of $W(0)$ as a function of the time $t$ spent after the memory has been prepared close to $\ket{C_\alpha^-}$. Note the much longer timescale for this single photon decay compared to the sub $\mu$s time needed to prepare Fock state $|1\rangle$ in (a). Dashed line: Fit of the exponential relaxation to vacuum.}
\label{fig:kappa_1_measurement}
\end{figure}

The measurement of $\kappa_1$ is done by observing the decay from Fock state $\ket{1}$ to the vacuum. If we prepare the state $\ket{C_\alpha^-}$ in the memory, and let it evolve under the action of two-photon dissipation at a rate $\kappa_2\gg \kappa_1$ (Fig.~\ref{fig:kappa_1_measurement}a), the parity of the memory is preserved so that the memory state ends up in the subspace generated by $\{ \ket{0}, \ket{1} \}$ with the same parity as $\ket{C_\alpha^-}$: that is the Fock state $\ket{1}$. All that is necessary to measure $\kappa_1$ is then to monitor the memory parity $\pi W(0)/2$ as it evolves towards $1$, here corresponding to the vacuum. 

In order to prepare $\ket{C_\alpha^-}$, we prepare $\ket{C_\alpha^+}$ as in Fig.~\ref{fig:fig2} and then apply a $Z$ gate. Even if the preparation and gate are not optimized, as in this measurement, the decay rate can still be extracted with excellent accuracy as it only affects the initial value of $W(0)$ during the decay from $\ket{1}$ to $\ket0$ (Fig.~\ref{fig:kappa_1_measurement}b).

The evolution of $W(0)$ in (Fig.~\ref{fig:kappa_1_measurement}b) is fitted by an exponential relaxation at a rate $\kappa_1/2\pi = 14~\mathrm{kHz}$. Repeating this measurement over the course of months, we found that it is not stable. The rate $\kappa_1/2\pi$ typically varies by $\pm 2~\mathrm{kHz}$ around this average value. 

\section{Calibration of the $Z$ gate}
\label{Appendix_Z_gate_calib}

\begin{figure}[!h]
\includegraphics[width=\linewidth]{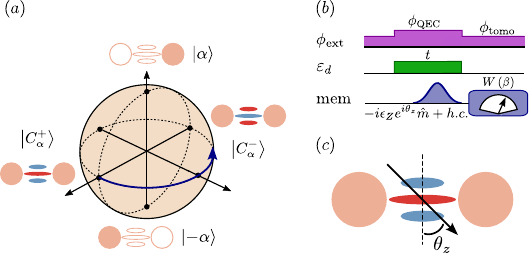}
\caption{(a) Bloch sphere of the cat qubit, whose computational basis is ${\ket{\alpha},\ket{-\alpha}}$. The effect of a $Z$ rotation is illustrated by the blue arrow. (b) Pulse sequence for $Z$ rotation characterization. (c) Effect of a drive $ -i \epsilon_Z e^{i \theta_z} \hat{m} + h.c.$ acting on the state $\ket{C_\alpha^+}$ of the memory.}
\label{fig:Presentation_Z_gate}
\end{figure}

The quantum operation shown in Fig.~\ref{fig:Zoscillations} consists of a rotation around the $z$ axis of the cat qubit Bloch sphere (Fig.~\ref{fig:Presentation_Z_gate}a). This is performed by driving the memory with a drive at frequency $\omega_m$, effectively implementing a displacement whose Hamiltonian reads 

\begin{equation*}
    \hat{H}_z / \hbar = -i \epsilon_Z e^{i \theta_z} \hat{m} + h.c..
\end{equation*}

\noindent Here, $\theta_z$ and $\epsilon_Z$ are the drive phase and amplitude. Applying this drive while simultaneously driving the buffer at $\phi_\mathrm{QEC}$ (Fig.~\ref{fig:Presentation_Z_gate}b) implements a quantum Zeno dynamics of the cavity which is restricted to states in the cat qubit subspace. The effect of $\hat{H}_z$ on the Wigner function of the memory then simply consists in shifting the phase of the interference fringes as seen in Fig.~\ref{fig:Presentation_Z_gate}c, effectively inducing the desired dynamics on the logical qubit.

\subsection{Calibration of the drive phase}

The first calibration needed for this scheme is to set the value of $\theta_z$ to $\pi$. This maximizes the gate speed for a given drive amplitude $\epsilon_Z$, improving the gate fidelity by decreasing the time during which single photon dissipation affects the memory. 

\begin{figure}[!h]
\includegraphics[width=\linewidth]{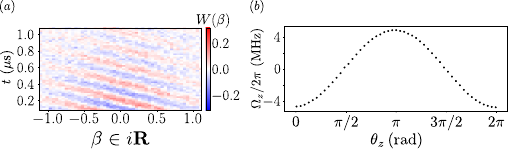}
\caption{(a) Measured Wigner function $W(\beta)$ of the memory as a function of $\beta\in i\mathbb{R}$ and time $t$. The displacement drive parameters are $\bar{\epsilon}_Z /2\pi = 1.25 \,\mathrm{MHz}$ and $\theta_z = \pi/2$ (b) Measured oscillation frequency $\Omega_z$ around the $z$ axis of the Bloch sphere as a function of $\theta_z$. }
\label{fig:Sup_mat_calib_Z_gate}
\end{figure}

This optimization is done by sweeping the phase of the memory drive and doing a vertical cut of the memory Wigner tomography. Looking at how fast the Wigner function fringes shift over time allows to extract the oscillation rate $\Omega_z$. Note that measuring $W\left(0 \right)$ alone would leave the sign of  $\Omega_z$ undetermined, which is why we measure a vertical cut of the Wigner function (Fig.~\ref{fig:Sup_mat_calib_Z_gate}a).

Doing this measurement for different value of $\theta_z$, (Fig.~\ref{fig:Sup_mat_calib_Z_gate}b) shows an evolution  $\Omega_z \left( \theta_z \right) \propto \cos\left(\theta_z \right)$. This is expected as only the vertical component of the drive $\mathrm{Re}\left(\epsilon_z e^{i\theta_z}\right)$ effectively displaces the fringes of the cat Wigner function (Fig.~\ref{fig:Zoscillations}c). The horizontal component $\mathrm{Im}\left(\epsilon_z e^{i\theta_z}\right)$ is disabled by the 2 photon dissipation and does not affect the system, which can be seen as a cancellation of $\Omega_z$ for $\theta_z = \pm\pi/2$.

\subsection{Comparison of $\kappa_\mathrm{Z}$ with the theoretical model}

\label{sec:kappaztheo}
The decay rate $\kappa_\mathrm{Z}$ of the oscillations around $Z$ (shown in Fig.~\ref{fig:Zoscillations}) is obtained by simulating the master equation
\begin{align}
    \label{eq:ME}
    \frac{d\hat \rho}{dt} =& -\frac{i}{\hbar}[\hat H , \hat \rho]\\
    &+ \mathcal{D}[\sqrt{\kappa_1}\hat m ](\hat \rho)+ \mathcal{D}[\sqrt{\kappa_{\varphi}^m}\hat m^\dagger \hat m](\hat \rho) \notag \\
    &+ \mathcal{D}[\sqrt{\kappa_b}\hat b](\hat \rho)+ \mathcal{D}[\sqrt{\kappa_{\varphi}^b}\hat b^\dagger \hat b](\hat \rho) \notag
\end{align}
where $\mathcal{D}[\hat{L}]\rho=\hat{L}\rho\hat{L}^\dagger-\hat{L}^\dagger \hat{L}\rho/2-\rho \hat{L}^\dagger \hat{L}/2$ is the Lindblad superoperator. The last four terms respectively model the single photon dissipation of the memory, the pure dephasing of the memory, the single photon dissipation of the buffer, and the pure dephasing of the buffer. The effective Hamiltonian of the system takes the form
\begin{align*}
    \frac{\hat{H}}{\hbar} = & \Delta_m \hat{m}^\dagger \hat{m} + \Delta_b \hat{b}^\dagger \hat{b} \\
    & - \frac{\chi_{m,m}}{2} \hat{m}^{\dagger 2} \hat{m}^2 - \frac{\chi_{b,b}}{2} \hat{b}^{\dagger 2} \hat{b}^2 - \chi_{m,b} \hat{m}^\dagger \hat{m} ~\hat{b}^\dagger \hat{b}\\
    & + g_2(\hat{m}^2 - \alpha^2)\hat{b}^\dagger + h.c\\
    & + i\epsilon_Z e^{-i\theta_z}\hat{m}^\dagger - i \epsilon_Z e^{i\theta_z}\hat m,
\end{align*}
where $\Delta_b = \omega_b - \omega_d$ is the detuning of the drive with respect to the buffer frequency, and  $\Delta_m = \omega_m -\omega_b/2 + \Delta_b/2$.
Here, the Gaussian drive envelope reads
\begin{align*}
   \epsilon_Z(t) = \bar{\epsilon}_Z\frac{6}{\sqrt{2\pi}}\exp{-\frac{(t-T/2)^2}{2w^2}},
\end{align*}

where the time window of the pulse is set to $w = T/6$, with $T$ being the total time of the pulse, and $\bar{\epsilon}_Z$ corresponding to the average drive amplitude. The case of a square pulse is easily extended by choosing $\epsilon_Z(t) = \bar{\epsilon}_Z$ over the same time window. As for the experiment, we fit the decaying oscillations of the photon number parity to extract the rotation frequencies $\Omega_Z$ and decay rate $\kappa_Z$ corresponding to each drive amplitude $\epsilon_Z$.

\begin{figure}[!h]
\includegraphics[width=\linewidth]{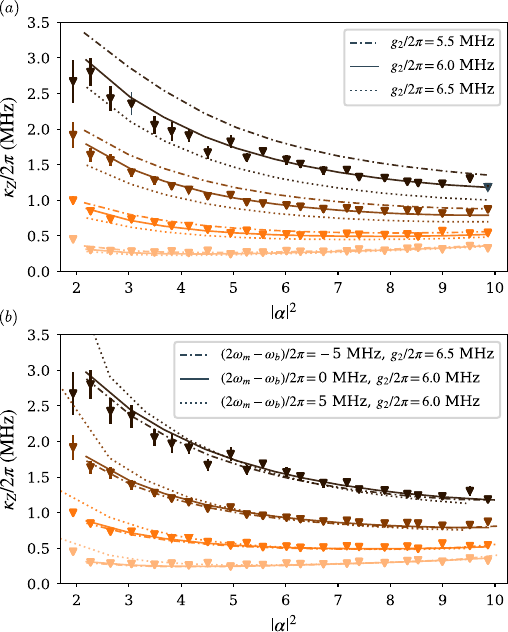}
\caption{(a) Triangles: measured decay rate $\kappa_Z$ of the oscillations around $Z$ as a function of $|\alpha|^2$ for four drive amplitudes $\epsilon_Z$ corresponding to distinct colors as in Fig.~\ref{fig:Zoscillations}d. Lines: simulated decay rates $\kappa_Z$ using Eq.~\eqref{eq:ME} with three values of the rate $g_2/2\pi$ indicated as an inset. (b) Triangles: same measurement as above. Lines: simulated $\kappa_Z$ for three values of the detuning between the buffer and the memory $(2\omega_m - \omega_b)/2\pi = -5,~0,~5~\mathrm{MHz}$ covering its the uncertainty range, and the corresponding optimal values of $g_2/2\pi = 6.5,~6.0,~6.0~\mathrm{MHz}$.}
\label{fig:Kappa_Z_vs_detuning}
\end{figure}

We numerically observe that the value of $\kappa_\mathrm{Z}$ is mostly sensitive to the loss rate $\kappa_1$, the ratio $4g_2^2/\kappa_b$, the self-Kerr rate $\chi_{m,m}$, and the effective detuning $\Delta_m$. We have independently measured $\chi_{m,m}/2\pi = 0.22~\mathrm{MHz}$, $\kappa_1/2\pi = 14~\mathrm{kHz}$, and $\kappa_b/2\pi = 40~\mathrm{MHz}$. 
This leaves us with two fit parameters, $g_2$ and $\Delta_m$. Let us estimate the uncertainty on $\Delta_m$. As we have measured $\Delta_b/2\pi = -3~\mathrm{MHz}$ (see Fig.~\ref{fig:Memory_01_Ramsey}), this leads to $\Delta_m/2\pi = \omega_m -\omega_b/2-1.5~\mathrm{MHz}$. The resonance condition $2\omega_m = \omega_b$ being sensitive to the external flux threading the loop, we estimate that $|2\omega_m - \omega_b|<5~\mathrm{MHz}$ from the width of the peak $\kappa_2(\varphi_\mathrm{ext})$ in Fig.~\ref{fig:fig2}a, leading to 
$\Delta_m/2\pi = -1.5 \pm 2.5 ~\mathrm{MHz}$. From the fit of the frequencies of the system and the formula derived in Sec.~\ref{Appendix:Hamiltonian_derivation}, we estimate $\chi_{m,b}^\mathrm{(model)}/2\pi \approx 1.6~\mathrm{MHz}$, $\chi_{b,b}^\mathrm{(model)}/2\pi \approx 10~\mathrm{MHz}$, and $g_2^\mathrm{(model)}/2\pi \approx 6.5~\mathrm{MHz}$. Interestingly, the decay rate $\kappa_Z$ strongly depends on $g_2$ and we use it to extract this parameter experimentally. For perfect frequency matching $2\omega_m = \omega_b$, the best fit to the simulation is obtained for $g_2/2\pi = 6\mathrm{MHz}$ (solid line in Fig.~\ref{fig:Kappa_Z_vs_detuning}).

To illustrate the sensitivity of the simulations to the value of $g_2$, we also compute $\kappa_Z(\alpha)$ for $g_2/2\pi=5.5\mathrm{MHz}$ and $g_2/2\pi=6.5\mathrm{MHz}$. The clear deviations in Fig.~\ref{fig:Kappa_Z_vs_detuning}a show that under the assumption that $2\omega_m = \omega_b$, $g_2$ can be determined with a much better precision than $2\pi\times 0.5 ~\mathrm{MHz}$ from the measured decay rates $\kappa_Z$.

The uncertainty on $g_2$ is actually dominated by the values it can take over the range of conceivable detunings $\Delta_m$. In order to get a higher bound on this uncertainty, we choose three values for the resonance condition: $(2\omega_m - \omega_b)/2\pi = -5,0,5~\mathrm{MHz}$, and search for the rate $g_2$ that best reproduces the experiment. The fitted $g_2$ rates are between $2\pi\times 6.0~\mathrm{MHz}$, and $2\pi\times 6.5~\mathrm{MHz}$ (Fig.~\ref{fig:Kappa_Z_vs_detuning}b). We therefore claim that $g_2/2\pi = 6\pm 0.5 \mathrm{MHz}$.

Note that in the simulations of the bit-flip time (Fig.~\ref{fig:figrates}) and in the simulations of the $Z$ gate (Fig.~\ref{fig:Zoscillations}), we have used the value $g_2/2\pi = 6~\mathrm{MHz}$ which best fits the gate rates and decay rates for the detuning $(2\omega_m - \omega_b)/2\pi =  3.5~\mathrm{MHz}$.

\subsection{Bias preserving nature of the $Z$ gate}

\label{appendix:bias_gate}
In order to preserve the benefit offered by bit-flip protection in cat qubits, it is crucial for logical gates to be bias-preserving~\cite{Puri2020}, meaning they do not convert phase-flip errors into bit-flip errors. 

\begin{figure}[!h]
\includegraphics[width=\linewidth]{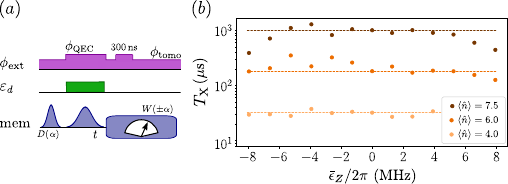}
\caption{(a) Pulse sequence for the $T_\mathrm{X}$ measurement, while continuously applying the $Z$ gate on the memory with an average drive amplitude $\bar{\epsilon}_Z$. (b) Dots: Measured $T_\mathrm{X}$ as a function of the average memory drive amplitude $\bar{\epsilon}_Z$ for different cat qubit sizes $\alpha=2$, $2.4$, and $2.7$ (yellow to brown). Lines: measured $T_\mathrm{X}$ at $\bar{\epsilon}_Z=0$.}
\label{fig:Sup_mat_bias_perserving}
\end{figure}

In order to verify the bias-preserving nature of the $Z$ gate, we measure the dependence of $T_\mathrm{X}$ when continuously driving the memory with a varying drive amplitude $\epsilon_Z$. Similarly to the measurement presented in Fig.~\ref{fig:figrates}, the flux is first set to $\phi_\mathrm{tomo}$ and the memory displaced by $\hat{D}\left(\alpha \right)$ in order to prepare the desired state $\ket{\alpha}$. The flux is then changed to $\phi_\mathrm{QEC}$ and 2 drives are sent to the buffer and memory modes, with amplitudes of $\epsilon_d \left( \alpha \right)$ and $\epsilon_Z$. The role of the drive acting on the buffer is to stabilize the cat qubit, preventing bit-flip errors from happening, while the drive acting on the memory performs the desired $Z$ gate. The Wigner function $W(\pm \alpha)$ is then finally measured for various waiting times, and their difference $W(\alpha)-W(-\alpha)\propto e^{-t/T_\mathrm{X}}$ is fitted to extract $T_\mathrm{X}$ (Fig.~\ref{fig:Sup_mat_bias_perserving}a). 

The measured dependence of $T_\mathrm{X}$ on $\bar{\epsilon}_Z$ is shown in Fig.~\ref{fig:Sup_mat_bias_perserving}b for different amplitudes $\alpha$. Despite a rather large measurement uncertainty, no visible decrease of $T_\mathrm{X}$ can be observed for $\left| \bar{\epsilon}_Z /2\pi \right| < 6\, \left(\mathrm{MHz} \right)$, after which the displacement becomes strong enough to overcome the stabilization provided by the two-photon dissipation. This induces an increased number of bit-flip errors, leading to a decrease of $T_\mathrm{X}$.

The $Z$ rotation presented in Fig.~\ref{fig:Zoscillations}a is measured with $\bar{\epsilon}_Z / 2\pi = 1.625 \, \mathrm{MHz}$. It can thus be assumed the drive did not induce additional bit-flip errors. Comparing the measured $T_\mathrm{X} \sim 10~\mathrm{ms}$ for $\alpha^2 = 9.3$ with the $Z\left(\pi \right)$ gate duration, we can estimate that bit-flip errors alone would then limit the gate fidelity to about $99.9997\%$. The obtained gate fidelity of $96.5\%$ thus primarily originates from phase-flip errors. 

\subsection{Fidelity of the $Z$ gate}

\label{sec:gatefidelity}
\begin{figure}[!h]
\includegraphics[width=\linewidth]{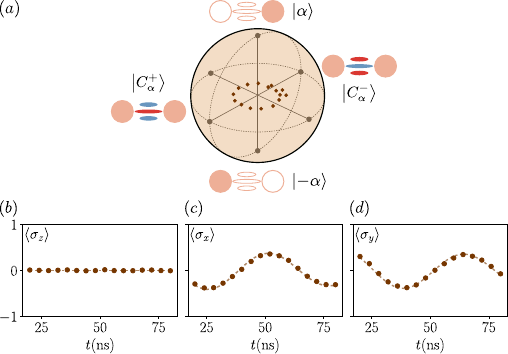}
\caption{(a) Trajectory of the cat qubit during the $Z$ gate estimated from the Wigner functions of Fig.~\ref{fig:Zoscillations}c. (b) Dots: Mean value of $\hat{\sigma}_{z, L}$ as a function of the gate time. Line: Linear fit of $\left< \hat{\sigma}_{z, L} \right>(t)$. (c) Dots: Mean value of $\hat{\sigma}_{x, L}$ as a function of the gate time. Line: fit used in Fig.~\ref{fig:Zoscillations}b to oscillations at a frequency $\Omega_Z/2\pi=19.8~\mathrm{MHz}$, decaying at a rate $\kappa_Z/2\pi=0.62~\mathrm{MHz}$. A scaling factor of $\pi/2$ is applied to respect the relation $W(0) = 2\left< \hat{\sigma}_{x, L} \right>/\pi$. (d) Dots: Mean value of $\hat{\sigma}_{y, L}$ as a function of the gate time. Line: fit used in (c) dephased by $\pi/2$.}
\label{fig:Sup_mat_Z_gate_fidelity_2D}
\end{figure}

To evaluate the fidelity of the $Z$ gate, we use the evolution of $W(\beta)$ shown in Fig.~\ref{fig:Zoscillations}c and directly estimate the evolution of the logical $\left< \hat{\sigma}_{x,L}\right>$, $\left< \hat{\sigma}_{y, L}\right>$ and $\left< \hat{\sigma}_{z, L}\right>$ from the Wigner functions. These operators are defined in the cat encoding as 

\begin{align*}
 & \hat{\sigma}_{x, L} = \ket{C_\alpha^+}\bra{C_\alpha^+} - \ket{C_\alpha^-}\bra{C_\alpha^-} \\
 & \hat{\sigma}_{y, L} = \ket{C_\alpha^{+i}}\bra{C_\alpha^{+i}} - \ket{C_\alpha^{-i}}\bra{C_\alpha^{-i}} \\
 & \hat{\sigma}_{z, L} = \ket{\alpha}\bra{\alpha} - \ket{-\alpha}\bra{-\alpha}, 
\end{align*}

\noindent with $\ket{C_\alpha^\pm} = \frac{\ket{\alpha} \pm \ket{-\alpha}}{\sqrt{2}}$ and $\ket{C_\alpha^{\pm i}} = \frac{\ket{\alpha} \pm i \ket{-\alpha}}{\sqrt{2}}$. The measured evolution of the three logical Bloch vector coordinates is shown in Fig.~\ref{fig:Sup_mat_Z_gate_fidelity_2D}a. Interestingly, the evolution of $\left< \hat{\sigma}_{z, L}\right>(t)$ during the gate shows no visible evolution (Fig.~\ref{fig:Sup_mat_Z_gate_fidelity_2D}b), which is expected from its bias preserving property Fig.~\ref{fig:Sup_mat_bias_perserving}b.

Using the formalism of the standard process matrix $\chi$~\cite{Nielsen2000}, the impact of the $Z$ gate on an initial density matrix $\hat{\rho}$ is modeled as 

\begin{equation}
\label{eq:standard_process_matrix}
    \mathcal{E}_Z\left(\hat{\rho} \right) = \sum_{m, n} \chi_{mn} \hat{E}_m \hat{\rho} \hat{E}_n^\dagger.
\end{equation}

\noindent where the fixed set of operators is chosen as  $\left\{\hat{E}_m \right\}_m \in \left\{\mathds{1}, \hat{\sigma}_{x, L}, \hat{\sigma}_{y, L}, \hat{\sigma}_{z, L}\right\}$. Furthermore, owing to the demonstrated negligible bit-flips occurring during the $Z$ gate, we assume no term causing a bit-flip type of error appears in Eq.~(\ref{eq:standard_process_matrix}) and only consider the simpler error model that provides a lower bound on the fidelity by neglecting the $\chi_{z,1}$ and $\chi_{1,z}$ terms
\begin{equation}
    \mathcal{E}_Z\left(\hat{\rho} \right) = \left(1-\varepsilon \right) \hat{\sigma}_{z, L} \hat{\rho} \hat{\sigma}_{z, L} + \varepsilon \hat{\rho}.
\end{equation}

\noindent The parameter $\chi_{z,z}=(1-\varepsilon)$ then corresponds to the gate fidelity owing to the definition $F = \mathrm{Tr}\left(\chi~ \chi_\mathrm{opt}\right)$. Indeed, the matrix $\chi_\mathrm{opt}$ is the $\chi$ matrix describing an ideal $Z$ gate, with all its terms null except for $\chi_{z,z} = 1$. Using the general form of a density matrix describing a qubit from its Bloch vector $(x~y~z)^T$

\begin{equation*}
    \hat{\rho} = \frac{1}{2}\left(\mathds{1} +  x\hat{\sigma}_{x, L}  + y\hat{\sigma}_{y, L} + z\hat{\sigma}_{z, L}\right),
\end{equation*}
the density matrix after application of the gate reads
\begin{equation*}
\begin{aligned}
    \mathcal{E}_Z\left(\hat{\rho} \right) = \frac{1}{2} &\left(\mathds{1} -  x\left(1 - 2\varepsilon \right)\hat{\sigma}_{x, L} \right. \\
    &\left. - y\left(1 - 2\varepsilon \right)\hat{\sigma}_{y, L} + \left< \hat{\sigma}_{z, L}\right>\hat{\sigma}_{z, L}\right).
\end{aligned}
\end{equation*}

\noindent The parameters $\varepsilon$ can then be estimated with the evolution of $\left< \hat{\sigma}_{x, L}\right>(t)$ and $\left< \hat{\sigma}_{y, L}\right>(t)$. We first check that both the evolution of $\left< \hat{\sigma}_{x, L} \right> = W(0) \pi/2$ and $\left< \hat{\sigma}_{y, L} \right>$ matches the fit of $W(0)$ used in Fig.~\ref{fig:Zoscillations}b and Fig.~\ref{fig:Zoscillations}c, up to a scaling factor and a dephasing. Finally, we compute $F=1/2+e^{-\pi\kappa_z/\Omega_z}/2=95\pm 2\%$ for the 26~ns long Gaussian pulse with $\left| \alpha \right|^2 = 9.3$ and for a drive amplitude $\bar{\epsilon}_Z/2\pi=1.625~\mathrm{MHz}$. The same expression for the measured decaying oscillations of parity in the case of square pulses leads to a slightly better fidelity of $F=96.5\pm 2\%$ in 28~ns. Note that the infidelity in the preparation of $\ket{C_\alpha^+}$ is due to a preparation time of $500$~ns starting from $\ket{0}$, similar to the phase-flip time of the cat qubit $1/\Gamma_\mathrm{Z}\approx 500$~ns for $\left| \alpha \right|^2 = 9.3$. This time should be optimized in future measurements. 

\section{Complementary data and analysis about bit-flip and phase-flip rates}
\label{Appendix:Simus}

\subsection{Measurement of $T_\mathrm{X}$, $\Gamma_\mathrm{Z}$}

\label{Appendix:Exemples_Tbf_Gammapf}

Figure \ref{fig:figrates} shows the measured $T_\mathrm{x}$ and $\Gamma_\mathrm{Z}$ for various cat sizes $\alpha$. An example of a phase-flip rate and bit-flip time measurement is shown in (Fig.~\ref{fig:TX_evo_vs_freq}) for $\alpha \approx 2.6$.

\begin{figure}[!h]
\includegraphics[width=\linewidth]{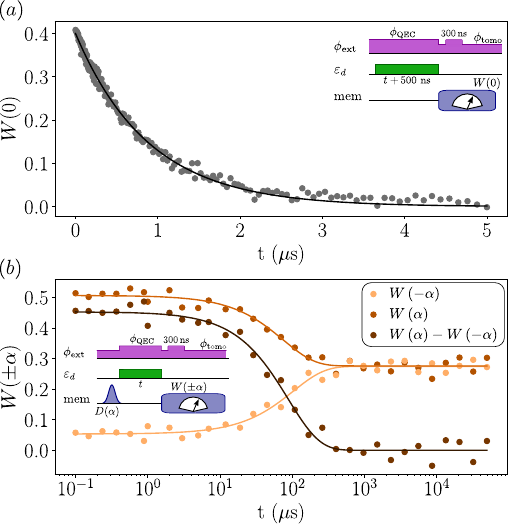}
\caption{Determination of $T_\mathrm{x}$ and $\Gamma_\mathrm{Z}$ for $\alpha^2 \approx 6.5$. (a) Dots: measured $W(0)$ as a function of cat stabilization time $t$. Solid line: fit with an exponential decay $e^{-\Gamma_\mathrm{z}t}$. Insert: associated pulse sequence. (b) Dots: measured $W(-\alpha)$, $W(\alpha)$ and $W(\alpha) - W(-\alpha)$ as a function of cat stabilization time $t$. Solid line: fit with an exponential decay of the measured data, the bit-flip time is deduced from the fit of $W(\alpha) - W(-\alpha)$. Insert: associated pulse sequence.}
\label{fig:Example_T_bf_Gamma_pf}
\end{figure}

The phase-flip rate corresponds to the loss of coherence of the cat qubit, through which any superposition of $\ket{\pm \alpha}$ decays to a mixture of these 2 coherent states. In order to probe it, a cat state $\ket{C_\alpha^+}$ is first prepared in the memory by applying a drive $\epsilon_d(\alpha)$ at $\phi_\mathrm{QEC}$, starting from an empty cavity. The cat qubit decoherence towards $\left(\ket{\alpha}\bra{\alpha} + \ket{-\alpha}\bra{-\alpha}\right) /2$ is then monitored by simply measuring $W(0)$. Fitting this evolution with an exponential decay at a rate $\Gamma_\mathrm{Z}$ gives the value of the phase-flip rate (Fig.~\ref{fig:Example_T_bf_Gamma_pf}a).

The bit-flip time $T_\mathrm{X}$ characterizes the typical time it takes for the populations in $\ket{\alpha}$ and $\ket{-\alpha}$ to equilibrate. We measure this value for various amplitudes $\alpha$ by first displacing the memory at $\phi_\mathrm{tomo}$ to prepare $\ket{\alpha}$, before applying a drive $\epsilon_d(\alpha)$ at $\phi_\mathrm{QE}$. The state $\ket{\alpha}$ is then protected by the Zeno dynamics. We monitor the values of $W(\alpha)$ and $W(-\alpha)$ over time, and fit $W(\alpha) - W(-\alpha)$ with an exponential decay at a rate $1/ T_\mathrm{X}$ which gives the value of the bit-flip time (Fig.~\ref{fig:Example_T_bf_Gamma_pf}b).

\subsection{Dependence of $T_\mathrm{X}$ and $\Gamma_\mathrm{Z}$ on photon number for various buffer drive frequencies}

The cat stabilization works by driving the buffer on resonance at $\phi_\mathrm{QEC}$. What happens if we drive it off resonant?

\begin{figure}[!h]
\includegraphics[width=\linewidth]{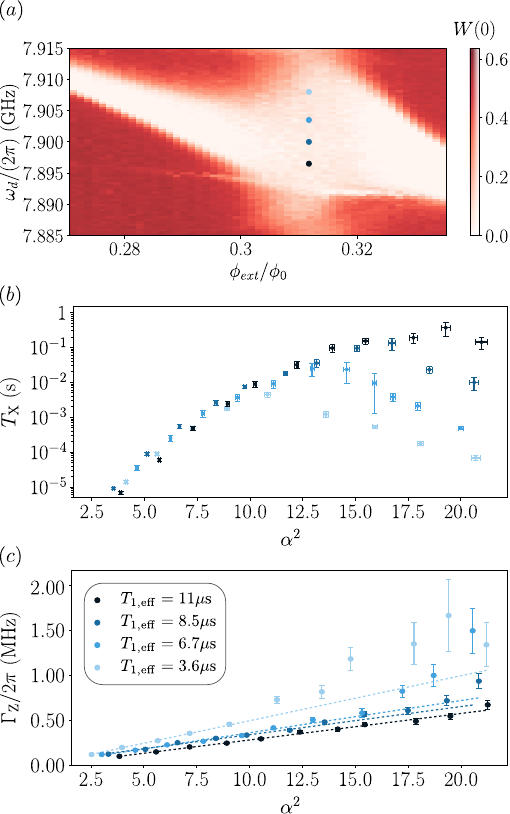}
\caption{(a) Anticrossing of the autoparametric memory-buffer system. Measured $W(0)$ after applying a buffer drive at $\omega_d$ for $5 \mu$s while at $\phi_\mathrm{ext}$. (b) Dots: measured $T_\mathrm{x}$ as a function of $\alpha^2$ for 4 different drive frequencies. The color matches with the dots shown in (a). (c) Dots: measured $\Gamma_\mathrm{Z}$ as a function of $\alpha^2$ for 4 different drive frequencies. Dashed line: Linear fit of the measured data with $\Gamma_\mathrm{Z} = 2 \left| \alpha \right|^2 / T_\mathrm{1, eff}$. where the fit parameter $T_\mathrm{1, eff}$ is shown as an inset.}
\label{fig:TX_evo_vs_freq}
\end{figure}

Figure \ref{fig:TX_evo_vs_freq}a shows the measured $W(0)$ as a function of flux and drive frequency $\omega_d$ after 5$\mu$s of stabilization. The red regions where $W(0)\approx 2/\pi$ correspond to a memory unaffected by the drive so that it is in the vacuum state. In contrast, a white region where $W(0)\ll 1$ corresponds to regions where a mixture of coherent states has formed in the memory. The figure reminds an avoided level crossing and it is actually an autoparametric version of that between $\omega_b(\phi_\mathrm{ext})$ and $2\omega_m(\phi_\mathrm{ext})$.

The dependence of $T_\mathrm{x}$ and $\Gamma_\mathrm{Z}$ on mean photon number $\alpha^2$ is measured for 4 buffer drive frequencies at $\phi_\mathrm{QEC}$ (dots in Fig.~\ref{fig:TX_evo_vs_freq}a). A similar behavior for $T_\mathrm{x}\left(\alpha^2 \right)$ can be seen for the 4 different drive frequencies, with an initial exponential increase before reaching a maximum for some optimal cat size (Fig.~\ref{fig:TX_evo_vs_freq}b). This optimal bit-flip time strongly depends on the chosen drive frequency, with the curve shown in Fig.~\ref{fig:figrates}d corresponding to the frequency that gives us the largest measured bit-flip time $T_\mathrm{X}$.

Note that the 1D-cuts of the Wigner functions $W(\beta)$ from which are extracted the bit-flip times (not shown here), exhibit a broadening of the Gaussian distribution around $\pm \alpha$ with increasing $\alpha$ and with increasing drive detuning. This can result from a distortion of the cat qubit manifold and/or a thermalization of the memory. As discussed in Sec.~\ref{Appendix:Bit_flip_time_limits}, based on reasonable assumptions, we show numerical evidence that both the smaller bit-flip times at other detuning and the broadening of the Wigner distribution can originate from the dephasing and the self-Kerr effect of the buffer mode.    

The dependence of $\Gamma_\mathrm{Z}\left(\alpha^2 \right)$ is measured for the same four drive frequencies (Fig.~\ref{fig:TX_evo_vs_freq}c). The same initial behavior can be seen for all 4 curves with an initial linear increase of the phase-flip rate. The slope of the linear increase is given by $\Gamma_\mathrm{Z} = 2 \left| \alpha \right|^2 / T_\mathrm{1, eff}$, with $T_\mathrm{1, eff}$ the effective memory lifetime. This effective lifetime matches the memory lifetime measured by looking at the decay from $\ket{1}$ to $\ket{0}$ at the optimal drive frequency (see section~\ref{kappa1determination}). The effective memory lifetime drastically deteriorates as the drive frequency detuning increases.

\subsection{Dephasing rate measurement}

\begin{figure}[!h]
\includegraphics[width=\linewidth]{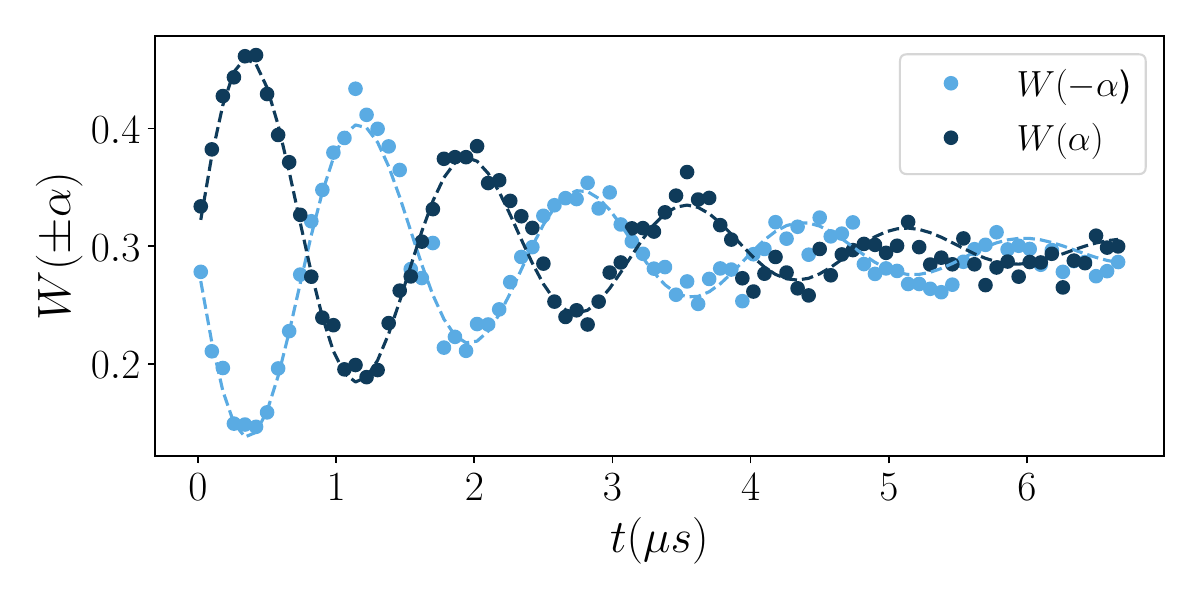}
\caption{Ramsey-like experiment on the $\left\{ \ket{0}, \ket{1}\right\}$ memory manifold. Dots: measured Wigner tomography $W(\pm \alpha)$. Dashed line: Fit used to extract the dephasing rate $\kappa_\varphi^{m}=2\Gamma_\varphi$.}
\label{fig:Memory_01_Ramsey}
\end{figure}

The memory pure dephasing rate $\kappa_\varphi^{m}$, used for numerical simulations of the bit-flip time evolution, is measured with a Ramsey-like experiment. A state close to $\left( \ket{0} + \ket{1} \right)/2$ is prepared in the memory by first displacing the memory to the coherent state $\ket{\alpha}$, with $\alpha = 2.1$, and letting it decay under the loss operator $\hat{L}_2 = \sqrt{\kappa_2}\hat{m}^2$. By Zeno effect, the 2-photon loss constraints the memory dynamics to the $\left\{ \ket{0}, \ket{1}\right\}$ manifold, hence acting as a qubit whose basis states are $0$ and $1$ photon in the memory. This state is then left idle for a time $t$, during which it rotates around the $Z$ axis of the Bloch sphere at the detuning $\Delta_m$ between the memory drive frequency and $\omega_\mathrm{m, QEC}$, in the frame of the drive frequency. The memory state evolves under the Hamiltonian and loss operator 
\begin{equation}
\begin{aligned}
    &\hat{H} / \hbar = -\frac{\Delta_m}{2} \left(\ket{1}\bra{1} - \ket{0}\bra{0}\right)\\
    &\hat{L}_1 = \sqrt{\kappa_1}\ket{0}\bra{1}, \, \hat{L}_\varphi = \sqrt{\frac{\Gamma_\varphi}{2}} \left(\ket{1}\bra{1} - \ket{0}\bra{0}\right).
\end{aligned}
\end{equation}
Note that the dephasing operator on the memory when Zeno blockade is disabled reads $\sqrt{\kappa_\varphi^{m}}\hat{m}^\dagger\hat{m}$. The correspondence is thus $\kappa_\varphi^{m}=2\Gamma_\varphi$.

For readout, the obtained state is then mapped into the $\left\{ \ket{\alpha}, \ket{-\alpha}\right\}$ manifold by driving the buffer mode with a drive $\epsilon_d\left(\alpha \right)$, mapping the eigenvectors of $\left(\ket{1}\bra{0} + \ket{0}\bra{1}\right)$ to the 2 coherent states. The Wigner function $W(\pm \alpha)$ (Fig.~\ref{fig:Memory_01_Ramsey}) is then measured, and the data fitted to extract the detuning $\Delta_m = 3~$MHz, and dephasing rate of the memory $\kappa_\varphi^{m}/2\pi \approx 0.16 \mathrm{MHz}$. 

\subsection{Population of the higher excited states of the transmon}

The transmon used for the Wigner tomography, dispersively coupled to the memory mode, has been shown to be one of the main factors limiting the bit-flip time $T_\mathrm{X}$ at large photon numbers. In this experiment, the 2 photon dissipation rate $\kappa_2$ is much greater than the dispersive shift $\kappa_2 \gg \chi$. This ensures that the population in the qubit first excited state only has a negligible detrimental impact on the cat qubit stabilization, and does not introduce additional bit-flip errors~\cite{Lescanne2020exponential, Berdou2022}. 

\begin{figure}[!h]
\includegraphics[width=8cm]{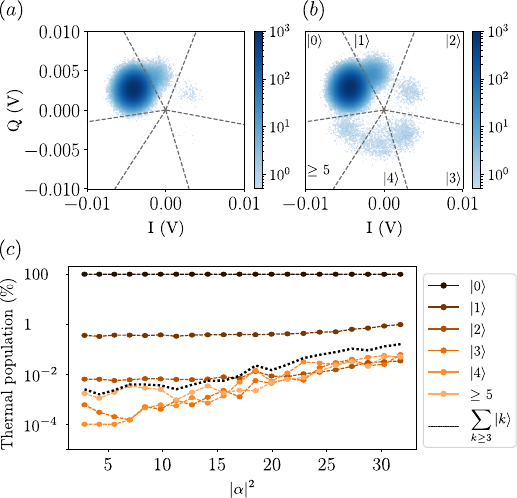}
\caption{(a) Histogram of $10^7$ measurements of the readout quadratures when the transmon and memory are in thermal equilibrium with the cold environment. (b) Same histogram when a cat qubit space is stabilized for $100~\mu\mathrm{s}$ with a mean photon number  $\left| \alpha \right|^2 \approx 30$. (c) Dots: measured occupation of the transmon states as a function of the stabilized mean number of photons in the memory. Excitation numbers are indicated by colors of increasing brightness. Black dotted line: sum of the populations in states higher than $|3\rangle$.}
\label{fig:Thermal_pop}
\end{figure}

However, simulations show that populations of higher excited states of the transmon have an impact on the bit-flip time. The transmon populations are probed by measuring the transmon state, while a cat qubit is stabilized in the memory mode. The drive frequency of the readout resonator is chosen in order to resolve the transmon states up to its $5^\mathrm{th}$ excited state. In contrast, the readout frequency used everywhere else in this work was optimized to distinguish between the transmon ground and first excited state. A cutoff is then calibrated to separate the states $\ket{0}, \ket{1}, \ket{2}, \ket{3}$, and $\ket{4}$ from the others, allowing to measure the transmon population for each amplitude $\alpha$ of the cat qubit (Fig.~\ref{fig:Thermal_pop}b).

As can be seen in Fig.~\ref{fig:Thermal_pop}c, the transmon populations in states of higher energy than $\ket{3}$ get excited for $\left| \alpha \right|^2 > 10$. This increase in transmon higher excitations is clearly correlated with the occupation of the memory mode. Besides it is not a Boltzmann distribution. 

We attribute the increase of the occupation of transmon states above $\ket{3}$  to a resonance between the memory and the higher levels of the transmon. The negative anharmonicity $\omega_{12}-\omega_{01} = - 2\pi \times 181~\mathrm{MHz}$ of the transmon results in a transition frequency between the 6-th and 7-th excited states being close to that of the memory. Note that such a resonance usually occurs when the resonator frequency is below that of the transmon. This phenomenon, which also happens with a simple Duffing oscillator, was recently investigated theoretically in Ref.~\cite{CohenPRX2023} and experimentally in Ref.~\cite{Khezri2022}.
As the number of photons in the resonator increases, the states below and above the 6-th and 7-th states of the transmon also hybridize.
In the steady state, we expect these hybridized states to be equally populated ~\cite{Ketzmerick2010,CohenPRX2023}, and this population to increase with the number of photons in the resonator. This qualitative signature of a growing number of hybridized states is observed in Fig.~\ref{fig:Thermal_pop}c. Note that this is in contrast with an overall increase in the temperature, where one would expect the hierarchy of populations to roughly follow the Boltzmann distribution. Below, we study how a finite population of these hybridized states sets an upper bound on the bit-flip time $T_\mathrm{X}$. 

\subsection{Impact of high excited states of the transmon on bit-flip time}

Populating the higher excited states of the transmon can in turn result in a shift of the memory frequency. If this shift exceeds the tolerance of the stabilization scheme, even a small population could limit a bit-flip time which is as high as hundreds of milliseconds. Below, we evaluate the magnitude of the frequency shifts that can be reached when the memory field drives the transmon.

In our system, the transmon is inductively coupled to the memory and capacitively coupled to a readout resonator. For simplicity, we neglect the Purcell filter of the readout resonator. The Hamiltonian of the system we consider reads
\begin{align}
\label{eq:Hamiltonian_static}
    \hat{H} =& 4E_C (\hat{n}_t-n_g)^2 - E_J \cos{\hat{\theta}_t} + \hbar \omega_m \hat{m}^\dagger \hat{m} + \hbar \omega_c \hat{c}^\dagger \hat{c}\notag \\
    & + \hbar g_{mt} \sin{\hat{\theta}_t}(\hat{m}+\hat{m}^\dagger) -  i \hbar g_{ct} n_t (\hat{c}-\hat{c}^\dagger),
\end{align}
where $\hat{n}_t$ and $\hat{\theta}_t$ are the transmon charge and phase operators, $n_g$ is the offset charge, $\hat{c}$ is the annihilation operator of the readout resonator, $\omega_c$ the frequency of the readout resonator, $g_{mt}/2\pi$ and $g_{ct}/2\pi$ are the coupling rates between the memory and the transmon, and between the readout resonator and the transmon.  We find the values of charging energy $E_C/h = 169.4~\mathrm{MHz}$ and Josephson energy $E_J/h = 22.85~\mathrm{GHz}$, as well as the values of the coupling rates $g_{ct}/2\pi = 67~\mathrm{MHz}$ and $g_{mt}/2\pi = 225~\mathrm{MHz}$, by fitting the measured low energy spectrum of the system which includes frequencies, anharmonicity and dispersive shifts of the system. Although we are concerned with the interaction of the memory and transmon, we included the readout resonator to correctly fit the spectrum of the system.

\begin{figure}
    \centering
    \includegraphics[width=\linewidth]{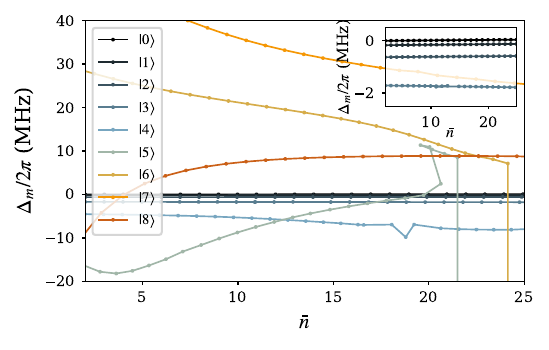}
    \caption{Dots: computed memory frequency shift as a function of mean photon number $\bar{n}$ in the memory for various transmon states. The vertical lines are due to state mistracking. Inset: zoom on the small dispersive shifts for low transmon excitation.}
    \label{fig:dispersive_shifts}
\end{figure}

By diagonalizing the Hamiltonian in Eq.~(\ref{eq:Hamiltonian_static}), one can obtain the values of the frequency shifts on the cavity as a function of the transmon state and the number of photons in the resonator (see Fig.~\ref{fig:dispersive_shifts}). In the inset, one recognizes the dispersive shift $\chi/2\pi = 0.170~\mathrm{MHz}$ when the transmon is in its first excited state $|1\rangle$, which is used for the Wigner tomography. While populating one of the first 3 excited states of the transmon cause small enough frequency shifts of the memory (inset) so that they are handled by the stabilization scheme, populating higher excited states can result in frequency shifts as large as $30~\mathrm{MHz}$. Such large frequency shifts are due to the non-perturbative hybridization of the transmon states, thus exiting the dispersive regime of the coupling between transmon and memory. 

\begin{figure}[!h]
    \centering
    \includegraphics[width=\linewidth]{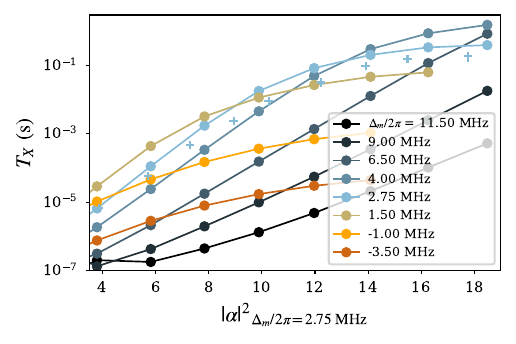}
    \caption{Effect of memory detuning $\Delta_m$ on the bit-flip time $T_X$. The bit-flip time is plotted as a function of the drive amplitude for several values of $\Delta_m$. The drive amplitude is converted to the corresponding photon number at $\Delta_m = 2.75~\mathrm{MHz}$, which approximately corresponds to the detunings set for Fig.~\ref{fig:figrates}b. The simulations are done under the assumption that $\kappa_\varphi^m/2\pi=0.16~\mathrm{MHz}$, $\kappa_\varphi^b=60\times \kappa_\varphi^m$, $\chi_{m,m}/2\pi=0.22~\mathrm{MHz}$, $\chi_{m,b}/2\pi=1.6~\mathrm{MHz}$ and without any self-Kerr on the buffer mode. The crosses indicate the experimentally measured bit-flip times shown in Fig.~\ref{fig:figrates}d.}
    \label{fig:Tx_vs_Delta}
\end{figure}

For small frequency shifts, the state remains confined to a manifold spanned by two coherent states, although the cat size might change slightly. Indeed, using a semi-classical analysis (Eq.~(S22) of~\cite{Berdou2022}), in the limit where $\kappa_1 \ll |\Delta_m|$ and  $\kappa_b \gg |\Delta_b|$, the photon number reads 
$|\alpha_\Delta|^2 = |\alpha_0|^2 - \frac{|\Delta_m|\kappa_b}{4|g_2|^2}$, where $|\alpha_0|^2$ is the photon number at zero detuning at the same drive amplitude. 
This equation has a solution if $|\Delta_m|< 4|g_2|^2|\alpha_0|^2/\kappa_b$. Using the measured and extracted values of $\kappa_b/2\pi = 40~\mathrm{MHz}$ and $g_2/2\pi = 6 ~\mathrm{MHz}$ , this leads to $|\Delta_m|<3.6 |\alpha_0|^2 ~\mathrm{MHz}$, that is the minimal condition for a 2D-manifold to be stabilized. The impact on the bit-flip time of the detuning $\Delta_m$ induced by the transmon higher excitation is illustrated in Fig.~\ref{fig:Tx_vs_Delta}, where the bit-flip time is plotted for various detunings as a function of the drive amplitude. The drive amplitude is converted to the photon number corresponding to the detuning $\Delta_m/2\pi = 2.75$ MHz. 

The detuning associated with the first and second excited states (below 20 photons) is less than $1~\mathrm{MHz}$, making the condition $|\Delta_m|<3.6 |\alpha_0|^2 ~\mathrm{MHz}$ largely satisfied. This is numerically verified in Fig.~\ref{fig:Tx_vs_Delta}. In this case, the bit flip rate becomes $T_\mathrm{X}^{-1} = \sum_{i,|\Delta_i|<\Delta_\mathrm{max}} p_i T_\mathrm{X}^{-1}(\Delta_i)$. The small detunings associated with the first and second excited states make the weighted contribution $p_i T_\mathrm{X}^{-1}(\Delta_i),~i=1,2$ negligible compared to $p_0 T_\mathrm{X}^{-1}(\Delta_0)$. 

However, detunings as large as $30~\mathrm{MHz}$ become difficult for the dissipation scheme to compensate.
For $|\Delta_m| = 30 ~\mathrm{MHz}$, the minimal condition to generate a cat state in the cavity is $ |\alpha_0|^2 > 8.5 $ photons. Moreover, even if the memory becomes populated, this large detuning comes with a large displacement on the buffer mode, $\beta \sim \Delta_m/2 g_2 = 2.5 $ (see next section).  Combined with its dephasing noise and large Kerr nonlinearity, the resulting bit-flip time of such a cat qubit is low. This is illustrated in Fig.~\ref{fig:Tx_vs_Delta}, where the bit-flip time corresponding to a detuning of $9~\mathrm{MHz}$ does not improve over the bare memory lifetime until $|\alpha_0|^2 = 12.5$. For simplicity, we assume that a bit-flip occurs every time the system is subject to such a large detuning for the range of photons considered here, which represents the worst-case scenario.  

From the analysis of the transmon, it is likely that populating the layer of hybridized states will result in a large detuning and therefore a bit-flip. This allows us to derive a simple upper bound on the induced bit-flip time $T_\mathrm{X}$ from the measured values of the state population: it is given by the inverse of the rate at which this layer of states is populated $\gamma_\mathrm{hyb}$.  Let us call $p_\mathrm{hyb}$ the population of the hybridized states, represented as a black dotted line in Fig.~\ref{fig:Thermal_pop}, and $p_1$ the population of the first excited state. Note that the state $\ket{2}$ merges with the rising plateau of states around $|\alpha|^2 = 20$ photons. The rate at which the hybridized layer gets populated thus reads $\gamma_\mathrm{hyb} = \gamma_\mathrm{hyb\rightarrow 1} p_\mathrm{hyb}$. Using the measured $\gamma_\mathrm{1\rightarrow 0}^{-1} = 18~\mu\mathrm{s}$ and assuming $\gamma_\mathrm{hyb\rightarrow 1}\approx\gamma_\mathrm{2\rightarrow 1} = 2\gamma_\mathrm{1\rightarrow 0}$, we obtain the red dotted line in Fig.~\ref{fig:figrates}b.

\subsection{Main limitation of the bit-flip time}
\label{Appendix:Bit_flip_time_limits}

Before studying other possible limiting factors on the bit-flip time, let us first briefly review the experimental data that cannot be explained solely by the heated transmon. As a reminder, the transmon higher excited state population may have been a limitation for bit-flip times above the red dots in Fig.~\ref{fig:figrates}d. However, measurements taken at various drive detunings reveal earlier saturation of the bit-flip time, almost two orders of magnitude lower than $0.3~\mathrm{s}$ (see Fig.~\ref{fig:TX_evo_vs_freq}b), which cannot be explained by the presence of a transmon. 

In this section,  we show that the self-Kerr effect of the memory steers the system away from resonance as the photon number increases, resulting in a smaller bit-flip time, which corroborates the dependence of the bit-flip time on the drive detuning.

Furthermore, we discuss possible mechanisms that could limit the bit-flip time even when the drive is on resonance ($\Delta_m=0$).

\subsubsection{Impact of memory self-Kerr on bit-flip time $T_\mathrm{X}$}

\begin{figure}[!h]
    \centering
    \includegraphics[width=0.97\linewidth]{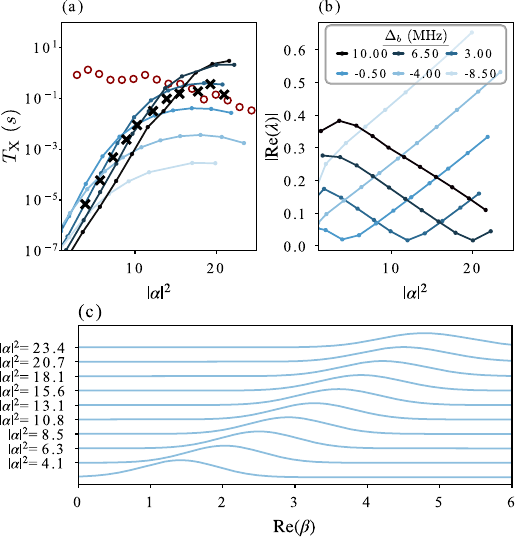}
    \caption{(a) Simulated bit-flip times for various values of the drive detuning $\Delta_b = \omega_b-\omega_d$ (color code as an inset in panel b), among which the values of Fig.~\ref{fig:TX_evo_vs_freq}b. The memory detuning from the buffer is $2\omega_m-\omega_b = 2\pi \times 3.5$ MHz. As a reference, we indicate the measured optimal bit-flip times (black crosses) and the bound independently set by the transmon (red circles). (b) Corresponding amplitudes of the buffer mode in the steady state as a function of average photon number $|\alpha|^2$ in the memory for the same detunings. (c) Simulated probability distributions of the quadrature defined by the direction of $\alpha$ in the memory phase space $\mathrm{P}(\mathrm{Re(\beta)}) = |\langle \mathrm{Re(\beta)}| \psi \rangle|^2$, plotted for several values of $|\alpha|^2$ (arbitrary units and offset for each $\alpha$). The detuning is set to $\Delta_b/2\pi = -4~\mathrm{MHz}$. As the photon number increases, the distribution broadens. All the simulations are done under the assumption that $\kappa_\varphi^m/2\pi=0.16~\mathrm{MHz}$, $\kappa_\varphi^b=60\times \kappa_\varphi^m$, $\chi_{m,m}/2\pi=0.22~\mathrm{MHz}$, $\chi_{m,b}/2\pi=1.6~\mathrm{MHz}$ and without any self-Kerr on the buffer mode. }
    \label{fig:Tx_vs_deltab_simulation}
\end{figure}

The protection against bit-flip is diminished away from resonance. The memory self-Kerr results in an effective detuning on the memory as the photon number increases. Each cat size corresponds to an optimal drive detuning for which the resulting effective detuning on the memory cancels. 
This is illustrated in Fig.~\ref{fig:Tx_vs_deltab_simulation}a, which shows the bit-flip times extracted from the simulation of Eq.~(\ref{eq:ME}), at the values of the drive detunings $\Delta_b$ corresponding to the experimental values of Fig.~\ref{fig:TX_evo_vs_freq}b and beyond. We observe a qualitatively similar behavior between simulations and measurement of the bit-flip times, in particular for the curves corresponding to larger detunings. 

Note that the self-Kerr rate of the buffer mode (not measured) is omitted here, as otherwise we observe that the required Hilbert space dimension for accurate enough simulations becomes prohibitively large for even moderate $|\alpha|$.

An effective detuning results in a displacement of the buffer mode, and computing this displacement can be used to estimate the effective detuning.
In the interaction picture, the master equation on the memory mode Eq.~(\ref{eq:ME}) reads 

\begin{align*}
    \frac{d\hat{m}}{dt} =& -i\Delta_m \hat{m} -i 2\chi_{m,m}\hat{m}^{\dagger}\hat{m}^2 -i\chi_{m,b}\hat{m}\hat{b}^\dagger \hat{b}\\  &- \frac{\kappa_\varphi^m+\kappa_1}{2}\hat{m} -i 2 g_2 \hat{m}^\dagger \hat{b}.
\end{align*}

Assuming a steady-state solution of the form $\rho = \ket{\alpha}\bra{\alpha}\otimes \ket{\lambda}\bra{\lambda}$, with $\alpha\neq 0$, and taking the trace of the above equation, we obtain
\begin{align*}
    \lambda = e^{i 2\theta_m}\frac{-\Delta_m-2\chi_{m,m}|\alpha|^2 - \chi_{m,b}|\lambda|^2+i(\kappa_\varphi^m+\kappa_1)/2}{2 g_2},
\end{align*}
where $\theta_m = \arg(\alpha)$.

In the limit $\kappa_\varphi^m, \kappa_1 \ll \left|\Delta_m +2\chi_{m,m}|\alpha|^2+\chi_{m,b}|\lambda|^2\right|$, the amplitude $\lambda$ becomes
\begin{align}\label{eq:displacement_detuning}
    \lambda = -e^{i 2\theta_m}\frac{\Delta_m+2\chi_{m,m}|\alpha|^2+ \chi_{m,b}|\lambda|^2}{2 g_2}.
\end{align}
The displacement amplitude on the buffer depends on the effective memory detuning, which in turn depends linearly on the cat photon number due to the self-Kerr effect. 
In Fig.~\ref{fig:Tx_vs_deltab_simulation}b are shown the amplitudes of the buffer mode corresponding to the curves of Fig.~\ref{fig:Tx_vs_deltab_simulation}a. For a given cat size, the drive detuning giving the optimal bit-flip time corresponds approximately to the smallest buffer amplitude and therefore to the smallest effective detuning.

\subsubsection{Impact of memory dephasing on bit-flip time $T_\mathrm{X}$}

\begin{figure}[!h]
    \centering
    \includegraphics[width=0.97\linewidth]{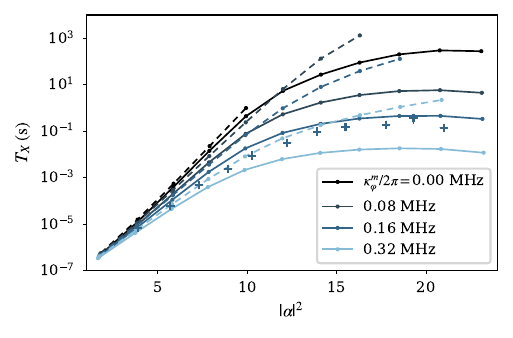}
    \caption{Simulated bit-flip times as a function of the memory dephasing rate $\kappa_\varphi^m$ for two values of the buffer dephasing rate, $\kappa_\varphi^b/2\pi = 0~\mathrm{MHz}$ (dashed lines) and $\kappa_\varphi^b/2\pi = 9.6~\mathrm{MHz}$ (solid lines). The simulations are done under the assumption that $\chi_{m,m}/2\pi=0.22~\mathrm{MHz}$, $\chi_{m,b}/2\pi=1.6~\mathrm{MHz}$ and without any self-Kerr on the buffer mode. }
    \label{fig:Tx_vs_kphis}
\end{figure}

We expect pure dephasing on the memory ($T_\varphi^m = 1/ \kappa_\varphi^{m}= 1~\mu\mathrm{s}$, see Fig.~\ref{fig:Memory_01_Ramsey}) to impact the bit-flip time $T_\mathrm{X}$. In the adiabatic regime where $8|\alpha|g_2 \ll \kappa_b$, the bit-flip time scales as $T_\varphi^m |\alpha|^{-2}\exp(2|\alpha|^2)$~\cite{Mirrahimi2014}. However, in our case, the adiabaticity criteria are not met for $|\alpha|\gtrsim 1$. 
As illustrated in Fig.~\ref{fig:Tx_vs_kphis}, the memory dephasing noise limits the scaling of $T_X$ with the photon number but does not lead to a saturation (dashed lines). 

Moreover, if one assumes the dephasing rates of the memory and buffer modes are limited by flux noise, we can estimate the buffer dephasing rate as
\begin{equation*}
    \kappa_\varphi^b = \kappa_\varphi^m \frac{\left|\frac{\partial \omega_b}{\partial\phi_\mathrm{ext}}(\phi_\mathrm{QEC})\right|}{\left|\frac{\partial \omega_m}{\partial\phi_\mathrm{ext}}(\phi_\mathrm{QEC})\right|} \approx  60 \kappa_\varphi^m.
\end{equation*}
Using the measured value $\kappa_\varphi^m/2\pi = 0.16~\mathrm{MHz}$, we estimate $\kappa_\varphi^b/2\pi = 9.6~\mathrm{MHz}$. When taking this large dephasing rate into account, the bit-flip time is predicted to saturate even without the transmon (solid lines in Fig.~\ref{fig:Tx_vs_kphis}).

\subsubsection{Impact of buffer thermalization on bit-flip time $T_\mathrm{X}$}

During the stabilization, the driven buffer ideally stays in the vacuum state. However, the coupling to the environment may lead to the thermalization of the buffer and memory bipartite system, which induces bit-flip errors. Indeed,
\begin{itemize}
    \item If the buffer state becomes thermal, buffer photons are converted to memory photons by the two-photon exchange term, thus creating a heating term of the form $\hat{m}^{\dagger 2}$ on the memory, thus affecting the bit-flip time $T_X$~\cite{Gautier2022}.
    \item The cross-Kerr effect ($\chi_{m,b}$) between memory and buffer modes leads to an effective dephasing rate of the memory. For instance, we estimate that with $0.2$ thermal photons in the buffer, the memory dephasing rate increases by $0.5~\mathrm{MHz}$~\cite{Clerk2007}.
\end{itemize}

We envision a few origins for the thermalization.
\begin{itemize}
    \item When the buffer is displaced to a finite amplitude $\lambda = \langle \hat b \rangle_\mathrm{ss} \neq 0$ due to drive detuning (see Fig.~\ref{fig:Tx_vs_deltab_simulation}b), dephasing noise on the buffer can be upconverted to thermal photons via the drive. Indeed, dephasing noise can be seen as small random rotations in phase space resulting in a small diffusion of a coherent state along the circle of radius $|\langle \hat b \rangle_\mathrm{ss}|$. This diffusion competes with single photon loss of the buffer, resulting in an effective temperature of $n_\mathrm{th}^b = \kappa_\varphi^b  |\langle \hat b \rangle_\mathrm{ss}|^2/\kappa_b$, where $\kappa_\varphi^b$ is the dephasing rate of the buffer. We find $n_\mathrm{th}^b = 0.24|\lambda|^2$. In Fig.~\ref{fig:Tx_vs_deltab_simulation}c, we show the marginals of the Wigner functions of the steady state in the memory as a function of $|\alpha|^2$. The memory heating can be seen as a broadening of the Gaussian distribution.
    \item The large expected self-Kerr rate of the buffer $\chi_{bb}/2\pi \approx 10~\mathrm{MHz}$ results in a small squeezing on the buffer when it is displaced. The single photon loss channel on a squeezed buffer yields an additional effective thermal occupation given by $n_\mathrm{th}^b = \sinh^2(r)$, where $r$ is the effective squeezing parameter. 
\end{itemize}

\section{Comparison between cat qubit stabilization schemes}
\label{Appendix:comparison}

To better understand the strengths and weaknesses of the autoparametric scheme compared to other experimental demonstrations, we show a comparison of key figures and constraints in Table~\ref{Tab:Comparison_different_designs}. In this table, $\epsilon_0$ is the dimensionless pump amplitude on the ATS (see (S3) in ~\cite{Lescanne2020exponential}). As can be seen in the table, the autoparametric cat performs better than other schemes on most figures, including the number of required microwave lines. The symmetry requirements are stronger than in other schemes if the buffer/flux line position is optimized to prevent memory leakage but do not preclude the feasibility of a repetition code based on this circuit. Besides, a notch filter on the drive line can be used to enhance or replace the protection provided by symmetry. We do not include the earlier scheme of a transmon qubit used as a coupler~\cite{Leghtas2015,Touzard2018} since no exponential scaling could be demonstrated owing to parasitic Hamiltonian terms.

The saturation due to the transmon higher excited states that we observe in $T_X$ could be canceled by using the technique introduced in Ref.\cite{Reglade2023} on an ATS where the readout transmon is removed. Finally, the main remaining caveat seems to be the buffer thermalization that limits the bit-flip scaling time in the autoparametric cat. We note that detuning the buffer drive by the right amount for a given cat qubit size should cancel out this effect (see Fig.\ref{fig:Tx_vs_deltab_simulation}).

\renewcommand{\arraystretch}{3}
\begin{table*}
\centering
\caption{State-of-the-art in key figures of the autoparametric cat, cats stabilized by an ATS, and Kerr-cats qubit. $^*$ The gate fidelity is here given for a $\pi/2$ gate instead of the $\pi$ gate in the main text in order to better compare with other works.\label{Table_key_figs}}
\begin{footnotesize}
\begin{tabular}{c|c|c|c}
\hline

\multicolumn{1}{c}{\textbf{Metric}} & \multicolumn{1}{c}{\textbf{Kerr cats}~\cite{Grimm2020, Frattini2022}} & \multicolumn{1}{c}{\textbf{ATS stabilized cat}~\cite{Lescanne2020exponential, Reglade2023}} & \multicolumn{1}{c}{\textbf{Autoparametric cat}}\\

\hline

$g_2$ & 
Irrelevant & 
$\begin{aligned}
    g_2 &= E_J \epsilon_0 \varphi_{\mathrm{ZPF}, b} \left(\varphi_{\mathrm{ZPF}, m} \right)^2 \\
    &= 2 \pi \times 0.76~\mathrm{MHz}
\end{aligned}$ &
$\begin{aligned}
    g_2 &= E_J \overbrace{\mathrm{sin}\left(\bar{\varphi}_J \right)}^{\gg \epsilon_0} \varphi_{\mathrm{ZPF}, b} \left(\varphi_{\mathrm{ZPF}, m} \right)^2 \\
    &= 2 \pi \times 6~\mathrm{MHz}
\end{aligned}$ \\

\hline

\makecell{Number of RF/DC/both \\ lines} & 2/1/0 & 3/0/2 & 1/0/1 \\

\hline

\makecell{$\hat{Z}\left(\pi/2 \right)$ gate time (ns) \\ $\frac{\pi}{2\mathrm{Re}(4\alpha \epsilon_Z)}$}  & 24~ns & 118~ns & 14~ns\\

\hline 

\makecell{$\hat{Z}\left(\pi/2 \right)$ gate fidelity \\ $F=1/2+e^{-\pi\kappa_z/2\Omega_z}/2 $} & $85.7 \%$ & $85.3 \%$ & $ 98 \%^*$ \\

\hline

CNOT gate& 
Theoretical proposal~\cite{Puri2020gates} &
\makecell{Theoretical proposal~\cite{Guillaud2019} $\&$ \\ experimental implementation on \\ coherent states~\cite{MM2023CNOT}} & 
See sec.~\ref{sec:cnot_gate} \\

\hline
$\kappa_2 / \kappa_1$ & Irrelevant & $0.9 \times 10^2$ & \makecell{Adiabatic elimination limits \\ it to $1.5 \times 10^2$} \\

\hline

Maximal $T_\mathrm{X}$  & \makecell{Up to about a ms then \\ observed decrease for larger $|\alpha|^2$ \\ likely due to non-RWA terms} & \makecell{Unknown \\ up to 10~s demonstrated} & \makecell{Up to $0.3~$s limited by excitation \\of transmon higher states \\ or buffer thermalization}\\

\hline

Geometrical constraint & Translation symmetry~\cite{Frattini2022} & Junction symmetry & Junction symmetry\\

\hline

Memory protection  & Bandpass filter & Notch filter & \makecell{In situ protection by symmetry,\\ notch filter is optional}\\

\hline

$T_\mathrm{X}$ scaling in photon number &  \makecell{Increase by successive steps \\ No clear observation\\ of exponential scaling yet} & \makecell{Exponential scaling \\ Limited by breakdown of the \\ adiabatic elimination} & \makecell{Exponential scaling \\ Limited by breakdown of the \\ adiabatic elimination \\ $\&$ buffer thermalization}\\

\hline
\end{tabular}
\end{footnotesize}
\label{Tab:Comparison_different_designs}
\end{table*}

\section{CNOT gate}
\label{sec:cnot_gate}
In order to correct the remaining phase flip errors of the autoparametric cat, one can use a chain of coupled autoparametric cats in order to perform phase flip error correction using a 1D repetition code. Such error correction code needs a two-qubit gate between cat qubits. Here we propose to use a CNOT gate and explain how it would operate.

\begin{figure}[!th]
    \centering
    \includegraphics[width=\linewidth]{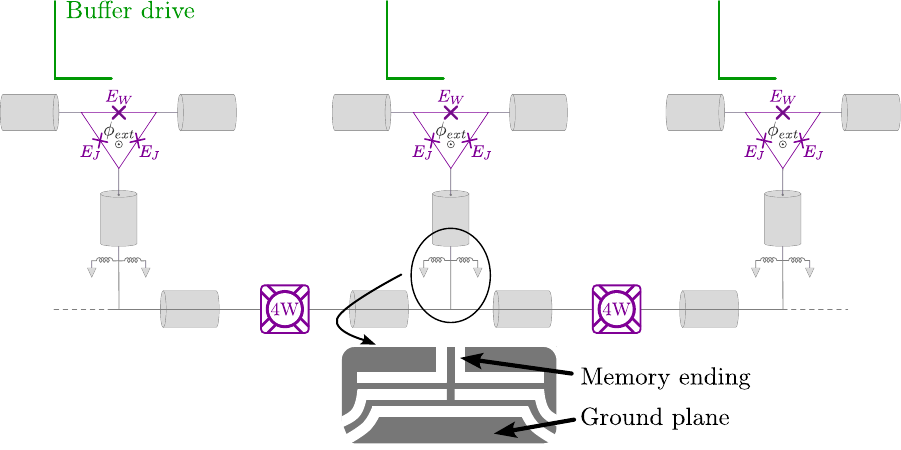}
    \caption{Possible layout for a repetition code using the autoparametric cat. The coupling between two neighboring cats is provided by a four-wave mixing element (such as a Josephson junction, SQUID, ATS or Quarton) driven at the frequency of the control cat to perform a CNOT gate.}
    \label{fig:scheme_couplage_chain}
\end{figure}

A CNOT gate can be performed between two autoparametric cats (named control and target) as long as they are coupled through a four-wave mixing element (Josephson junction, SQUID, ATS, quarton~\cite{Ye2020},\ldots, see Sec.\ref{sec:cnot_H}). A scheme for a repetition code coupling autoparametric cats with 4-wave mixers is shown in Fig.~\ref{fig:scheme_couplage_chain}.

A CNOT gate between two cat qubits should act on coherent states as
\begin{align*}
    \hat{U}_\mathrm{CNOT} \ket{\alpha_c, \pm \alpha_t} = \ket{\alpha_c, \pm \alpha_t}, \\
    \hat{U}_\mathrm{CNOT} \ket{-\alpha_c, \pm \alpha_t} = \ket{-\alpha_c, \mp \alpha_t},
\end{align*}

\noindent where $\alpha_c$ and $\alpha_t$ are the control and coherent state amplitudes of the target cat qubit. Thus the CNOT gate can be viewed as a $\pi$ rotation of the target coherent state conditioned on the control state. However, an interesting feature of bosonic code, key to implement bias preserving gates, is the ability to redefine the logical basis so that the computational basis before and after a gate can differ~\cite{Guillaud2019, Albert2016}. Using this feature, one can define a CNOT gate as
\begin{align}
    \label{eq:cnot_gate_eq1}
    \hat{U}_\mathrm{CNOT} \ket{\alpha_c, \pm \alpha_t} = \ket{\alpha_c, \pm \alpha_t e^{i \pi/2}}, \\
    \hat{U}_\mathrm{CNOT} \ket{-\alpha_c, \pm \alpha_t} = \ket{-\alpha_c, \pm \alpha_t e^{-i \pi/2}},
    \label{eq:cnot_gate_eq2}
\end{align}
with the following change of basis $\{\ket{\alpha_t}, \ket{-\alpha_t}\} \rightarrow \{\ket{i \alpha_t}, \ket{-i\alpha_t}\}$ for the target cat code. The CNOT then becomes a rotation by $\pm \pi/2$ of the target coherent state conditioned on the control state. If the target cat stabilization is turned off, such a gate can be performed using the following Hamiltonian
\begin{equation*}
    \hat{H} = \hbar g_\mathrm{CNOT} (\ket{\alpha_c}\bra{\alpha_c} - \ket{-\alpha_c}\bra{-\alpha_c}) \hat{a}_t^\dagger \hat{a}_t
\end{equation*}
and can be approximated by 
\begin{equation}
    \hat{H} =\hbar g_\mathrm{CNOT} (e^{i\phi_c} \hat{a}_c^\dagger + e^{-i\phi_c} \hat{a}_c) \hat{a}_t^\dagger \hat{a}_t
    \label{eq:cnot_approx_H}
\end{equation}
where $\hat{a}_c$ and $\hat{a}_t$ are annihilation operators of the target and control memory and $\phi_c = \arg(\alpha_c)$.

\subsection{CNOT Hamiltonian engineering}
\label{sec:cnot_H}
The Hamiltonian we propose for the CNOT gate is generated using a four-wave mixing interaction coming from a non-linear coupler. One can write this interaction as
\begin{align}
    \label{eq:cnot_ATS_H}
    \hat{H}=&\hbar g_4 (\xi(t) + \phi_c (\hat{a}_c^\dag + \hat{a}_c) + \phi_t (\hat{a}_t^\dag + \hat{a}_t))^4, 
\end{align}
where $g_4$ is the four-wave mixing interaction strength, 
$\phi_c$ (respectively $\phi_t$) the phase zero point fluctuations of the control (respectively target) memory across the non-linear coupler, $\hat{a}_c$ (respectively $\hat{a}_t$) the annihilation operator of the control (respectively target) memory and $\xi(t)$ the amplitude of an RF drive on the non-linear coupler.

Using a drive  amplitude $\xi(t) = \xi_0 \cos(\omega_c t)$ at the frequency of the control memory $\omega_c$  and
in the frame rotating at the control memory frequency $\omega_c$, the Hamiltonian (\ref{eq:cnot_ATS_H} ) under rotating wave approximation leads to the interaction

\begin{align}
    \label{eq:cnot_ATS_H_RWA}
    \hat{H} = 12 \hbar g_4 \phi_c \phi_t^2 (\xi_0 \hat{a}_c^\dag  + \xi_0^\star \hat{a}_c) \hat{a}_t^\dag \hat{a}_t,
\end{align}

which can be written as:
\begin{align}
    \label{eq:cnot_H}
    \hat{H}_\mathrm{CNOT} = \hbar g_\mathrm{CNOT}(e^{i\arg(\xi_0)} \hat{a}_c^\dag  + e^{-i\arg(\xi_0)} \hat{a}_c)  \hat{a}_t^\dag \hat{a}_t,
\end{align}
with $g_\mathrm{CNOT} = 12 g_4 \phi_c \phi_t^2 |\xi_0|$. To get the Hamiltonian of Eq.~(\ref{eq:cnot_approx_H}), one has to tune the phase of $\xi_0$ such that $\arg(\xi_0) = \arg(\alpha_c)$. We note that the larger two-photon coupling rate $g_2$ of the autoparametric cat would permit to increase the drive amplitude $\xi_0$ without affecting the stabilization of the control cat qubit. Ultimately, we expect larger gate fidelities and lower gate times owing to the autoparametric cats.

\subsection{CNOT sequence}
The CNOT gate described by Eqs.~(\ref{eq:cnot_gate_eq1}) and  (\ref{eq:cnot_gate_eq2}) can be performed with the following sequence
\begin{enumerate}
    \item Turn off the stabilization of the target cat. One can do it by turning off the buffer drive and moving away from the $\phi_\mathrm{QEC}$ flux point.
    \item Turn on the CNOT Hamiltonian Eq.~(\ref{eq:cnot_approx_H}) by driving the nonlinear coupler at $\omega_c$ during a time $T_g = \pi/(4 |\alpha_c| g_\mathrm{CNOT})$.
    \item Turn back on the stabilization of the target with a buffer drive phase shifted by $\pi$ such that the stabilized sub-space is $\{\ket{i \alpha_t}, \ket{-i\alpha_t}\}$. The stabilization has to be kept during a time larger than $1/|\alpha_t|^2\kappa_{2,t}$, where $\kappa_{2,t}$ is the target two-photon dissipation rate, such that the target memory state is projected onto the cat qubit manifold $\{\ket{i \alpha_t}, \ket{-i\alpha_t}\}$.
\end{enumerate}

\renewcommand{\arraystretch}{1.3}

\begin{table*}[!th]
\centering
\caption{\textbf{Estimated parameters of the device and the associated measurement method.}}
\begin{tabular}{c|c|c}
\hline
\hline
\multicolumn{1}{c}{\textbf{Parameter}} & \multicolumn{1}{c}{\textbf{Value}} & \multicolumn{1}{c}{\textbf{Method of determination}}\\
\hline
2 photon dissipation flux $\phi_\mathrm{2 ph}$ & 0.312 $\phi_0$  & Memory and buffer spectroscopies \\

Tomography flux $\phi_\mathrm{tomo}$ & 0.168 $\phi_0$  & Optimization of the memory displacements $\hat{\mathcal{D}}\left( \beta \right)$\\

Sweet spot $\phi_\mathrm{ext}^\mathrm{(sweet)}$ & 0.4 $\phi_0$  & Memory and buffer spectroscopies\\

Memory frequency $\omega_m\left(\phi_\mathrm{2 ph} \right)/2\pi$ & $3.948$~GHz & Memory and buffer spectroscopies\\

Buffer frequency $\omega_b\left(\phi_\mathrm{2 ph} \right)/2\pi$ & $7.896$~GHz & Memory and buffer spectroscopies\\

Transmon frequency $\omega_\mathrm{q}/2\pi$ & $5.387$~GHz & Ramsey interferometry at $\phi_\mathrm{tomo}$\\

Readout resonator frequency $\omega_\mathrm{r}/2\pi$ & $6.967$~GHz & Readout spectroscopy\\

Effective inductive energy $\bar{E}_J\left(\phi_\mathrm{2 ph} \right)/\hbar$ & $228$~GHz & Memory and buffer spectroscopies\\

Effective inductive energy $\bar{E}_W\left(\phi_\mathrm{2 ph} \right)/\hbar$ & $51$~GHz & Memory and buffer spectroscopies\\

Effective inductive energy $\bar{E}_J\left(\phi_\mathrm{tomo} \right)/\hbar$ & $242$~GHz & Memory and buffer spectroscopies\\

Effective inductive energy $\bar{E}_W\left(\phi_\mathrm{tomo} \right)/\hbar$ & $97$~GHz & Memory and buffer spectroscopies\\

Memory participation ratio $\varphi_{\mathrm{ZPF}, m}$ & 0.0305 & Memory and buffer spectroscopies\\

Buffer participation ratio $\varphi_{\mathrm{ZPF}, b}$ & 0.0648 & Memory and buffer spectroscopies\\

Predicted two-to-one coupling rate $g_2/2\pi$ & $6.2~\mathrm{MHz}$ & Memory and buffer spectroscopies\\

\hline

Memory self-Kerr $\chi_\mathrm{m, m}$ & $0.220$~MHz & Coherent state deformation\\

Buffer self-Kerr $\chi_\mathrm{b, b}$ & $10$~MHz & Estimated using Appendix B\\

Transmon to memory cross-Kerr $\chi_\mathrm{q, m}$ & $0.170$~MHz & Ramsey interferometry with a displaced cavity\\

Buffer to memory cross-Kerr $\chi_\mathrm{b, m}$ & $1.6$~MHz & Estimated using Appendix B\\

Transmon to readout cross-Kerr $\chi_\mathrm{q, r}$ & $3.5$~MHz & Readout spectroscopy with excited transmon\\

Two-to-one coupling rate $g_2/2\pi$ & $6 \pm 0.5~$MHz  & Fidelity of logical $\hat{Z}$ rotations\\

\hline

Memory single photon loss $\kappa_1/2\pi$ & $14$~kHz & Decay $\ket{1} \rightarrow \ket{0}$\\

Memory effective 2 photons loss $\kappa_2/2\pi$ & $2.16$~MHz & Decay $\ket{C_\alpha^+} \rightarrow \ket{0}$ or $\ket{C_\alpha^-} \rightarrow \ket{1}$ \\

Memory dephasing rate $\kappa_\varphi^m/2\pi$ & $0.08$~MHz & Ramsey interferometry Zeno blocked on $\mathrm{span}\left(\ket{0}, \ket{1} \right)$\\

Buffer single photon loss $\kappa_b/2\pi$ & $40$~MHz & Buffer spectroscopy\\

Readout resonator linewidth $\kappa_r/2\pi$ & $1.8$MHz & Readout spectroscopy\\

Transmon $T_1$ & $18~\mu \mathrm{s}$ & Standard decay measurement \\

Transmon $T_2$ & $15~\mu \mathrm{s}$ & Ramsey interferometry\\

Memory thermal population $n_\mathrm{th, m}$ & $0.011$ photon & Transmon Ramsey interferometry\\

Transmon thermal population $n_\mathrm{th, q}$ & $0.015$ photon & Standard transmon readout\\

\hline
\hline
\end{tabular}
\label{Tab:Device_parameters}
\end{table*}

\clearpage

\end{document}